\documentclass[a4paper,11pt]{article}
\pdfoutput=1 \usepackage{jheppub_2} 
\usepackage{booktabs}
\usepackage[T1]{fontenc} 
\usepackage{graphicx, multirow,soul,url,amsmath,amsfonts,amssymb,mathrsfs,amsfonts}
\usepackage[all]{hypcap}
\usepackage{url}
\usepackage{bbm}
\usepackage{changepage}
\usepackage{cancel,ulem}
\usepackage{sidecap}
\usepackage{graphicx}
\usepackage{slashed}
\usepackage[dvipsnames]{xcolor}
\usepackage{multicol, blindtext}
\usepackage[section]{placeins}
\usepackage[version=3]{mhchem}
\usepackage{caption}
\expandafter\def\csname ver@subfig.sty\endcsname{}
\usepackage{svg}
\usepackage{lipsum}
\usepackage{hyperref}
\usepackage{footnote}
\usepackage{bbold}
\usepackage{amsbsy}
\usepackage[export]{adjustbox}
\usepackage{units}

\newcommand{\figref}[1]{Fig.~\ref{#1}}

\newcommand{\secref}[1]{Section~\ref{#1}}

\usepackage{float}
\usepackage{pifont}
\restylefloat{table}

\def\thefootnote{\fnsymbol{footnote}}
\allowdisplaybreaks

\newcommand{\cmark}{\ding{51}}%
\newcommand{\xmark}{\ding{55}}%

\def\chain#1#2{\mathrel{\mathop{\null\longrightarrow}\limits^{#1}_{#2}}}
\newcommand{\1}{{\bf 1}}
\newcommand{\2}{{\bf 2}}
\newcommand{\3}{{\bf 3}}
\newcommand{\4}{{\bf 4}}

\newcommand{\bq}{\begin{eqnarray}}
\newcommand{\nq}{\end{eqnarray}}

\usepackage{subfig}

\title{\bf Confronting SO(10) GUTs with proton decay and
gravitational waves}
\author[a]{Stephen F. King,}
\affiliation[a]{School of Physics and Astronomy, University of Southampton, Southampton, SO17 1BJ, U.K.}
\author[b,c]{Silvia Pascoli,}
\affiliation[b]{Dipartimento di Fisica e Astronomia, Universit\`a di Bologna e INFN, Sezione di Bologna, via Irnerio 46, I-40126 Bologna, Italy}
\author[c]{Jessica Turner}
\affiliation[c]{Institute for Particle Physics Phenomenology, Department of Physics, Durham University, Durham, U.K.}
\author[d,e]{and Ye-Ling Zhou}
\affiliation[d]{School of Fundamental Physics and Mathematical Sciences, Hangzhou Institute for Advanced
Study, UCAS, Hangzhou, China}
\affiliation[e]{International Centre for Theoretical Physics Asia-Pacific, Beijing/Hangzhou, China}

\emailAdd{s.f.king@soton.ac.uk}
\emailAdd{silvia.pascoli@unibo.it}
\emailAdd{jessica.turner@durham.ac.uk}
\emailAdd{zhouyeling@ucas.ac.cn}

\abstract{Grand Unified Theories (GUT) predict proton decay as well as the formation of 
cosmic strings which can generate gravitational waves.
We determine which non-supersymmetric $SO(10)$ breaking chains provide gauge unification in addition to a gravitational signal from cosmic strings. We calculate the GUT and intermediate scales for these $SO(10)$ breaking chains by solving the renormalisation group equations at the two-loop level.
 This analysis predicts the GUT scale, hence the
proton lifetime, in addition to the scale of cosmic string generation and thus the associated gravitational wave signal.
We determine which $SO(10)$ breaking chains survive in the
event of the null results of the next generation of gravitational waves and proton decay searches and determine the correlations between proton decay and gravitational waves scales if these observables are measured.  }

\preprint{
\begin{flushleft}{
IPPP/20/120}
\end{flushleft}
}


\begin{document}

\thispagestyle{empty}
\def\thefootnote{\fnsymbol{footnote}}
\setcounter{footnote}{1}

\setcounter{page}{0}
\maketitle
\vspace{-1cm}
\flushbottom

\def\thefootnote{\arabic{footnote}}
\setcounter{footnote}{0}

\section{Introduction}
Grand Unified Theories (GUTs) \cite{Georgi:1974sy} are frameworks that aim to unify the strong, weak, and electromagnetic interactions of the 
Standard Model (SM) into a single gauge group, $G_{X}$, at scale, $M_{X},$ with associated coupling,
$g_X$. The $SO(10)$ \cite{Fritzsch:1974nn,Georgi:SO10} gauge group is one of the most well studied GUT symmetries 
since it makes neutrino mass inevitable and also provides unification without the need for supersymmetry \cite{Bertolini:2009qj,Bertolini:2009es,Chakrabortty:2017mgi,Chakrabortty:2019fov,Meloni:2019jcf,Ohlsson:2020rjc}. 
 At scales lower than $M_X$, the group gets broken into its subgroups and ultimately down to the SM gauge group via a Higgs mechanism. The breaking from $SO(10)$ to the SM gauge group can proceed via different intermediate symmetry groups, which depend on the choice of the extended Higgs sector. We refer to this pattern of breaking as a ``breaking chain''. The main prediction of GUTs is proton decay.
While this process has not yet been observed, there are stringent constraints on the proton lifetime \cite{Miura:2016krn,Abe:2013lua,Abe:2014mwa,Heeck:2019kgr} and, therefore, the scale of grand unification, 
$M_{X}$. From SuperKamiokande (Super-K), the most robust existing constraint places many of the breaking chains of $SO(10)$ under significant tension. 
Moreover, future experiments, such as Hyper-Kamiokande (Hyper-K) \cite{Abe:2018uyc}, DUNE \cite{Acciarri:2016crz}, and JUNO \cite{An:2015jdp}, are expected to improve the sensitivity to this process by up to one order of magnitude. As a large fraction of theoretical GUT models predicts proton lifetimes below the $10^{36}$ years, these experiments are getting tantalisingly close to the potential discovery of this process and the groundbreaking result of the baryon number violation.

In addition to proton decay, another generic prediction of GUTs is the production of topological defects, which occurs when the GUT symmetry is spontaneously broken down to the SM gauge group, which often occurs in a series of intermediate steps. 
The presence of certain defects, such as monopoles and domain walls, is problematic as these would come to dominate the Universe's energy density. This problem is solved by advocating a period of inflation after their production to suppress their density strongly.

Cosmic strings are produced if a $U(1)$ gauge subgroup of the GUT is broken and arise in most GUT symmetries \cite{Jeannerot:2003qv}\footnote{ In some cases, larger subgroups, as such as $SU(2)$ can also lead to cosmic strings.}. The cosmic string network exhibits a scaling solution
and therefore does not overclose the Universe. This network can survive and generate a source of gravitational radiation \cite{Vilenkin:1984ib,Caldwell:1991jj,Hindmarsh:1994re}. The possibility of using gravitational waves (GW) generated by cosmic strings
to probe GUT scale physics has been recently explored \cite{Dror:2019syi,Buchmuller:2019gfy,Chakrabortty:2020otp}. 

In our recent paper \cite{King:2020hyd},
we demonstrated the non-trivial complementarity between the observation of
proton decay and gravitational waves in assessing the viability of $SO(10)$ GUT breaking chains. 
We showed that future non-observations could exclude $SO(10)$
breaking via flipped $SU(5)\times U(1)$ or standard $SU(5)$,
while breaking via a Pati-Salam intermediate symmetry \cite{Pati:1974yy},
or standard $SU(5)\times U(1)$, may
be favoured. Further, we highlighted that recent results by the NANOGrav experiment \cite{Arzoumanian:2018saf} can be interpreted 
in such frameworks as an indication of cosmic strings at a scale $\sim 10^{12}$~GeV. 

In this paper, we expand on the methodology outlined in our previous work \cite{King:2020hyd} and present a detailed and systematic study of the proton decay and gravitational wave predictions of all non-supersymmetric 
$SO(10)$ breaking chains.
We highlight the importance of gauge coupling unification in constraining the GUT and intermediate breaking scales, and consequently, the scale of cosmic string generation and the proton decay rate.
We then examine the viability of these chains in light of the future experimental landscape. There are 31 possible ways to break $SO(10)$ to the SM gauge group ($G_{\rm SM}$), which provide gauge unification and a possible GW signal. We compute the proton decay lifetime for each breaking chain by performing a renormalisation group equation (RGE) analysis at the two-loop level to find the GUT scale and intermediate symmetry breaking scale. This RGE analysis also provides the scale of cosmic string generation and, therefore, a prediction of the gravitational wave spectrum  generated from the string network. Importantly, this work provides the correlation between proton decay and gravitational waves of all $SO(10)$ breaking chains and
detailing which chains will potentially be excluded by null observation in both sets of experiments.
In our analysis, we impose the following criteria:
\begin{itemize}
\item[1)] We systematically include all symmetry breaking chains of $SO(10)$ GUTs. 
\item[2)] We include the minimal particle content consistent with the SM and neutrino masses as observed by neutrino oscillation experiments. This means that, in the fermion sector, only the SM fermions and right-handed neutrinos will be considered, and in the Higgs sector, only those Higgses used to generate fermion masses and achieve the GUT symmetry breaking will be present. 
\item[3)] We do not include effects that may induce additional mass scales beyond those associated with the pattern of GUT breaking. Namely, we do not include supersymmetric, or threshold effects \cite{Weinberg:1980wa,Hall:1980kf}. The former introduces the supersymmetry breaking scale, and the latter may induce significant radiative corrections from heavy particle masses. These effects may alter the conclusions of our analysis, and their study is deferred to future works. 
\item[4)] We approximate the cosmic strings as Nambu-Goto strings such that the gravitational radiation is the primary source of energy emission of the cosmic string network.
\item[5)] We assume that cosmic strings evolution occurs in the standard radiation- and matter-dominated eras. Inflation is assumed to inflate away other problematic topological defects, such as domain walls and monopoles generated from the GUT breaking. We assume inflation completes before the formation of the string network. This will happen when the scales of the breaking inducing cosmic strings and other defects are well separated, and inflation takes place in between. Only in these scenarios, corresponding to specific breaking chains, observable GW from cosmic string scaling can arise. Such a scenario is the focus of this paper. 
\end{itemize}

The remainder of the paper is organised as follows: in \secref{sec:classify} we present our classification of all $SO(10)$ breaking chains and introduce the Higgs multiplets used to break the various intermediate symmetries. We highlight the 31 chains that are testable through a combination of proton decay and GW experiments.  
 We follow in \secref{sec:unificationRGE} with a discussion of our methodology of the RGE analysis. In \secref{sec:PDresults} and \secref{sec:GW1} 
 we provide a detailed discussion of how we calculate the proton lifetime and GW signal for each testable breaking chain, respectively. We discuss the interplay between proton decay and GWs in \secref{sec:stringtension} and 
 finally we summarise and conclude in \secref{sec:sumanddis}.
 
\section{The SO(10) GUT framework}\label{sec:classify}
$SO(10)$ is the minimal Grand Unified gauge symmetry which predicts proton decay and could generate an observable gravitational signature of cosmic strings. 
We begin with a brief review of the $SO(10)$ GUT framework, including its breaking chains to the SM gauge symmetry and topological defects. 

$SO(10)$ can be broken into the SM gauge group in various ways. 
Each ``breaking chain'' has a distinct pattern of intermediate gauge symmetries and we use the following abbreviations for these
gauge groups:
\begin{eqnarray} \label{eq:symmetries}
&&G_{\rm SM} = SU(3)_c \times SU(2)_L \times U(1)_Y\,, \nonumber\\
&&G_{51} = SU(5) \times U(1)_V \,, \nonumber\\
&&G_{51}^{\rm flip} = SU(5)^{\rm flip} \times U(1)_V \,, \nonumber\\
&&G_{3221} = SU(3)_c \times SU(2)_L \times SU(2)_R \times U(1)_X \,, \nonumber\\
&&G_{3211} = SU(3)_c \times SU(2)_L \times U(1)_R \times U(1)_X \,, \nonumber\\
&&G_{421} = SU(4)_c \times SU(2)_L \times U(1)_R\,, \nonumber\\
&&G_{422} = SU(4)_c \times SU(2)_L \times SU(2)_R \,,
\end{eqnarray}
where $G_{422}$ is the Pati-Salam gauge group and the charge $X$ is correlated with $B-L$ via $X = \sqrt{\frac{3}{2}}\frac{B-L}{2}$. The breaking of $SO(10)$ can include an intrinsic $Z_2^C$ parity symmetry. In the Pati-Salam model, this parity symmetry represents the interchange of $(\overline{\4}, \2, \1)$ and $(\4,\1,\2)$. 
We abbreviate the Pati-Salam gauge group which preserves this ``left-right'' symmetry as $G_{422}^C \equiv G_{422} \times Z_2^C$. The left-right symmetry may also be preserved in $G_{3221}$, and we denote this as $G_{3221}^C \equiv G_{3221} \times Z_2^C$.  
All breaking chains from $SO(10)$ to $G_{\rm SM}$ are listed in Table~\ref{tab:chains_1} and \ref{tab:chains_2}. 
In our previous paper \cite{King:2020hyd}, we classified these chains into four categories:
\begin{itemize}
\item[(a)] breaking chains with the standard $SU(5)$ and a $U(1)$ as an intermediate symmetry.
\item[(b)] breaking chains with the flipped $SU(5)\times U(1)$ as intermediate symmetry. 
\item[(c)] breaking chains with the Pati-Salam symmetry $G_{422}$ or its subgroups as intermediate symmetry.
\item[(d)] breaking chains with the standard $SU(5)$ subgroup as the lowest intermediate scale before breaking to $G_{\rm SM}$. 
\end{itemize}
Table.~\ref{tab:chains_1} lists all breaking chains of types (a), (b) and (d). Table~\ref{tab:chains_2} contains the 31 breaking chains of type (c), which includes 6, 12, 10 and 3 chains with one (I), two (II), three (III), and four (IV) intermediate symmetries. As types (a), (b) and (d) cannot achieve unification without supersymmetry, we focus exclusively on type (c) breaking chains as they can be tested through a combination of proton decay and gravitational wave detection.

\begin{table}[t!]
\centering
\begin{tabular}{|p{10mm} p{6mm} p{6mm}p{6mm} p{6mm}c|}
\hline
\multirow{2}{*}{$SO(10)$} & \multirow{2}{*}{$\chain{\rm defect}{\rm Higgs}$} & \multirow{2}{*}{$G_1$} & \multirow{2}{*}{$\chain{\rm defect}{\rm Higgs}$} & \multirow{2}{*}{$G_{\rm SM}$} & Observable \\
& & & & & strings? \\\hline
aI: & $\chain{\rm m}{\bf 45}$ & $G_{51}$ & $\chain{\rm m,s}{\overline{\bf 126},{\bf 45}}$ & & \xmark
\\
bI: & $\chain{\rm m}{\bf 45}$ & $G_{51}^{\rm flip}$ & $\chain{\rm s}{\overline{\bf 126},{\bf 45}}$ & & \cmark
\\
dI: & $\chain{}{\bf 16}$ & $SU(5)$ & $\chain{\rm m}{\overline{\bf 126}}$ & & \xmark
\\[2mm]
\hline
\end{tabular} \\[2mm]
\begin{tabular}{|p{10mm} p{6mm} p{6mm}p{6mm}p{6mm}p{6mm} p{6mm}c|}
\hline
\multirow{2}{*}{$SO(10)$} & \multirow{2}{*}{$\chain{\rm defect}{\rm Higgs}$} & \multirow{2}{*}{$G_2$} & \multirow{2}{*}{$\chain{\rm defect}{\rm Higgs}$} & \multirow{2}{*}{$G_1$} & \multirow{2}{*}{$\chain{\rm defect}{\rm Higgs}$} & \multirow{2}{*}{$G_{\rm SM}$} & Observable \\
& & & & & & & strings? \\\hline
aII: & $\chain{\rm m}{\bf 45}$ & $G_{51}$ & $\chain{\rm m}{\bf 45}$ & $G_{3211} $ & $\chain{\rm s}{\overline{\bf 126}}$ & & \cmark
\\
dII: & $\chain{\rm m}{\bf 45}$ & $G_{51}$ & $\chain{\rm s}{\bf 16}$ & $SU(5) $ & $\chain{\rm m}{\overline{\bf 126}}$ & & \xmark 
\\[2mm]
\hline
\end{tabular}
\caption{Type (a), (b), (d) breaking chains with one (I) and two (II) intermediate gauge symmetries. The representation of the Higgs multiplet responsible for the spontaneous symmetry breaking is listed under the arrow. Topological defects, including monopoles (m), strings (s) and domain walls (w), induced by such breaking, are listed above arrows. A blank above an arrow implies that no defects are formed. An explanation of observable strings are given in the main text.}\label{tab:chains_1}
\end{table}
\begin{table}[t!]
\centering
\begin{tabular}{|p{10mm} p{6mm} p{6mm}p{6mm} p{6mm}c|}
\hline
\multirow{2}{*}{$SO(10)$} & \multirow{2}{*}{$\chain{\rm defect}{\rm Higgs}$} & \multirow{2}{*}{$G_1$} & \multirow{2}{*}{$\chain{\rm defect}{\rm Higgs}$} & \multirow{2}{*}{$G_{\rm SM}$} & Observable \\
& & & & & strings? \\\hline
I1: & $\chain{\rm m}{\bf 45}$ & $G_{3221} $ & $\chain{\rm s}{\overline{\bf 126}}$ & & \cmark
\\
I2: & $\chain{\rm m,s}{\bf 210}$ & $G_{3221}^C $ & $\chain{\rm s,w}{\overline{\bf 126}}$ & & \xmark
\\
I3: & $\chain{\rm m}{\bf 45}$ & $G_{421} $ & $\chain{\rm s}{\overline{\bf 126}}$ & & \cmark
\\
I4: & $\chain{\rm m}{\bf 210}$ & $G_{422} $ & $\chain{\rm m}{\overline{\bf 126},{\bf 45}}$ & & \xmark
\\
I5: & $\chain{\rm m,s}{\bf 54}$ & $G_{422}^C $ & $\chain{\rm m,w}{\overline{\bf 126},{\bf 45}}$ & & \xmark
\\
I6: & $\chain{\rm m}{\bf 210}$ & $G_{3211} $ & $\chain{\rm s}{\overline{\bf 126}}$ & & \cmark
\\[2mm]
\hline
\end{tabular}
\\[2mm]
\begin{tabular}{|p{10mm} p{6mm} p{6mm}p{6mm}p{6mm}p{6mm} p{6mm}c|}
\hline
\multirow{2}{*}{$SO(10)$} & \multirow{2}{*}{$\chain{\rm defect}{\rm Higgs}$} & \multirow{2}{*}{$G_2$} & \multirow{2}{*}{$\chain{\rm defect}{\rm Higgs}$} & \multirow{2}{*}{$G_1$} & \multirow{2}{*}{$\chain{\rm defect}{\rm Higgs}$} & \multirow{2}{*}{$G_{\rm SM}$} & Observable \\
& & & & & & & strings? \\\hline
II1: & $\chain{\rm m}{\bf 210}$ & $G_{422}$ & $\chain{\rm m}{\bf 45}$ & $G_{3221} $ & $\chain{\rm s}{\overline{\bf 126}}$ & & \cmark
\\
II2: & $\chain{\rm m,s}{\bf 54}$ & $G_{422}^C$ & $\chain{\rm m}{\bf 210}$ & $G_{3221}^C $ & $\chain{\rm s,w}{\overline{\bf 126}}$ & & \xmark
\\
II3: & $\chain{\rm m,s}{\bf 54}$ & $G_{422}^C$ & $\chain{\rm m,w}{\bf 45}$ & $G_{3221} $ & $\chain{\rm s}{\overline{\bf 126}}$ & & \cmark
\\
II4: & $\chain{\rm m,s}{\bf 210}$ & $G_{3221}^C$ & $\chain{\rm w}{\bf 45}$ & $G_{3221} $ & $\chain{\rm s}{\overline{\bf 126}}$ & & \cmark
\\
II5: & $\chain{\rm m}{\bf 210}$ & $G_{422}$ & $\chain{\rm m}{\bf 45}$ & $G_{421} $ & $\chain{\rm s}{\overline{\bf 126}}$ & & \cmark
\\
II6: & $\chain{\rm m,s}{\bf 54}$ & $G_{422}^C$ & $\chain{\rm m}{\bf 45}$ & $G_{421} $ & $\chain{\rm s}{\overline{\bf 126}}$ & & \cmark
\\
II7: & $\chain{\rm m,s}{\bf 54}$ & $G_{422}^C$ & $\chain{\rm w}{\bf 210}$ & $G_{422} $ & $\chain{\rm m}{\overline{\bf 126},{\bf 45}}$ & & \xmark
\\
II8: & $\chain{\rm m}{\bf 45}$ & $G_{3221}$ & $\chain{\rm m}{\bf 45}$ & $G_{3211} $ & $\chain{\rm s}{\overline{\bf 126}}$ & & \cmark
\\
II9: & $\chain{\rm m,s}{\bf 210}$ & $G_{3221}^C$ & $\chain{\rm m,w}{\bf 45}$ & $G_{3211} $ & $\chain{\rm s}{\overline{\bf 126}}$ & & \cmark
\\
II10: & $\chain{\rm m}{\bf 210}$ & $G_{422}$ & $\chain{\rm m}{\bf 210}$ & $G_{3211} $ & $\chain{\rm s}{\overline{\bf 126}}$ & & \cmark
\\
II11: & $\chain{\rm m,s}{\bf 54}$ & $G_{422}^C$ & $\chain{\rm m,w}{\bf 210}$ & $G_{3211} $ & $\chain{\rm s}{\overline{\bf 126}}$ & & \cmark
\\
II12: & $\chain{\rm m}{\bf 45}$ & $G_{421}$ & $\chain{\rm m}{\bf 45}$ & $G_{3211} $ & $\chain{\rm s}{\overline{\bf 126}}$ & & \cmark
\\[2mm]
\hline
\end{tabular}
\\[2mm]
\begin{tabular}{|p{10mm} p{6mm} p{6mm}p{6mm}p{6mm}p{6mm}p{6mm}p{6mm} p{6mm}c|}
\hline
\multirow{2}{*}{$SO(10)$} & \multirow{2}{*}{$\chain{\rm defect}{\rm Higgs}$} & \multirow{2}{*}{$G_3$} & \multirow{2}{*}{$\chain{\rm defect}{\rm Higgs}$} & \multirow{2}{*}{$G_2$} & \multirow{2}{*}{$\chain{\rm defect}{\rm Higgs}$} & \multirow{2}{*}{$G_1$} & \multirow{2}{*}{$\chain{\rm defect}{\rm Higgs}$} & \multirow{2}{*}{$G_{\rm SM}$} & Observable \\
& & & & & & & & & strings? \\\hline
III1: & $\chain{\rm m,s}{\bf 54}$ & $G_{422}^C$ & $\chain{\rm w}{\bf 210}$ & $G_{422}$ & $\chain{\rm m}{\bf 45}$ & $G_{421}$ & $\chain{\rm s}{\overline{\bf 126}}$ & & \cmark
\\[2mm]
\hline
 \multicolumn{10}{|l|}{Continue on the next page}
 \\\hline
\end{tabular}
\caption{Type (c) breaking chains with one (I), two (II), three (III) and four (IV) intermediate gauge symmetries. Representations of Higgs fields that are key to achieve the breakings are listed under arrows. Topological defects, including monopoles (m), strings (s) and domain walls (w), induced by the breaking, are listed above arrows. }\label{tab:chains_2}
\end{table}
\begin{table}[t!]
\centering
\begin{tabular}{|p{10mm} p{6mm} p{6mm}p{6mm}p{6mm}p{6mm}p{6mm}p{6mm} p{6mm}c|}
\hline
 \multicolumn{10}{|l|}{Table~\ref{tab:chains_2} (Continued)}
 \\\hline
\multirow{2}{*}{$SO(10)$} & \multirow{2}{*}{$\chain{\rm defect}{\rm Higgs}$} & \multirow{2}{*}{$G_3$} & \multirow{2}{*}{$\chain{\rm defect}{\rm Higgs}$} & \multirow{2}{*}{$G_2$} & \multirow{2}{*}{$\chain{\rm defect}{\rm Higgs}$} & \multirow{2}{*}{$G_1$} & \multirow{2}{*}{$\chain{\rm defect}{\rm Higgs}$} & \multirow{2}{*}{$G_{\rm SM}$} & Observable \\
& & & & & & & & & strings? \\\hline
III2: & $\chain{\rm m,s}{\bf 54}$ & $G_{422}^C$ & $\chain{\rm w}{\bf 210}$ & $G_{422}$ & $\chain{\rm m}{\bf 45}$ & $G_{3221}$ & $\chain{\rm s}{\overline{\bf 126}}$ & & \cmark
\\
III3: & $\chain{\rm m,s}{\bf 54}$ & $G_{422}^C$ & $\chain{\rm w}{\bf 210}$ & $G_{422}$ & $\chain{\rm m}{\bf 210}$ & $G_{3211} $ & $\chain{\rm s}{\overline{\bf 126}}$ & & \cmark
\\
III4: & $\chain{\rm m,s}{\bf 54}$ & $G_{422}^C$ & $\chain{\rm m}{\bf 210}$ & $G_{3221}^C$ & $\chain{\rm w}{\bf 45}$ & $G_{3221}$ & $\chain{\rm s}{\overline{\bf 126}}$ & & \cmark
\\
III5: & $\chain{\rm m,s}{\bf 54}$ & $G_{422}^C$ & $\chain{\rm m}{\bf 210}$ & $G_{3221}^C$ & $\chain{\rm m,w}{\bf 45}$ & $G_{3211}$ & $\chain{\rm s}{\overline{\bf 126}}$ & & \cmark
\\
III6: & $\chain{\rm m,s}{\bf 54}$ & $G_{422}^C$ & $\chain{\rm m,w}{\bf 45}$ & $G_{3221}$ & $\chain{\rm m}{\bf 45}$ & $G_{3211} $ & $\chain{\rm s}{\overline{\bf 126}}$ & & \cmark
\\
III7: & $\chain{\rm m,s}{\bf 210}$ & $G_{3221}^C$ & $\chain{\rm w}{\bf 45}$ & $G_{3221}$ & $\chain{\rm m}{\bf 45}$ & $G_{3211} $ & $\chain{\rm s}{\overline{\bf 126}}$ & & \cmark
\\
III8: & $\chain{\rm m}{\bf 210}$ & $G_{422}$ & $\chain{\rm m}{\bf 45}$ & $G_{3221}$ & $\chain{\rm m}{\bf 45}$ & $G_{3211} $ & $\chain{\rm s}{\overline{\bf 126}}$ & & \cmark
\\
III9: & $\chain{\rm m,s}{\bf 54}$ & $G_{422}^C$ & $\chain{\rm m}{\bf 45}$ & $G_{421}$ & $\chain{\rm m}{\bf 45}$ & $G_{3211} $ & $\chain{\rm s}{\overline{\bf 126}}$ & & \cmark
\\
III10: & $\chain{\rm m}{\bf 210}$ & $G_{422}$ & $\chain{\rm m}{\bf 45}$ & $G_{421}$ & $\chain{\rm m}{\bf 45}$ & $G_{3211} $ & $\chain{\rm s}{\overline{\bf 126}}$ & & \cmark
\\[2mm]
\hline
\end{tabular}
\\[2mm]
\begin{tabular}{|p{10mm} p{6mm} p{6mm}p{6mm}p{6mm}p{6mm}p{6mm}p{6mm}p{6mm}p{6mm} p{6mm}c|}
\hline
\multirow{2}{*}{$SO(10)$} & \multirow{2}{*}{$\chain{\rm defect}{\rm Higgs}$} & \multirow{2}{*}{$G_4$} & \multirow{2}{*}{$\chain{\rm defect}{\rm Higgs}$} & \multirow{2}{*}{$G_3$} & \multirow{2}{*}{$\chain{\rm defect}{\rm Higgs}$} & \multirow{2}{*}{$G_2$} & \multirow{2}{*}{$\chain{\rm defect}{\rm Higgs}$} & \multirow{2}{*}{$G_1$} & \multirow{2}{*}{$\chain{\rm defect}{\rm Higgs}$} & \multirow{2}{*}{$G_{\rm SM}$} & Observable \\
& & & & & & & & & & & strings? \\\hline
IV1: & $\chain{\rm m,s}{\bf 54}$ & $G_{422}^C$ & $\chain{\rm m}{\bf 210}$ & $G_{3221}^C$ & $\chain{\rm w}{\bf 45}$ & $G_{3221}$ & $\chain{\rm m}{\bf 45}$ & $G_{3211}$ & $\chain{\rm s}{\overline{\bf 126}}$ & & \cmark
\\
IV2: & $\chain{\rm m,s}{\bf 54}$ & $G_{422}^C$ & $\chain{\rm w}{\bf 210}$ & $G_{422}$ & $\chain{\rm m}{\bf 45}$ & $G_{3221}$ & $\chain{\rm m}{\bf 45}$ & $G_{3211} $ & $\chain{\rm s}{\overline{\bf 126}}$ & & \cmark
\\
IV3: & $\chain{\rm m,s}{\bf 54}$ & $G_{422}^C$ & $\chain{\rm w}{\bf 210}$ & $G_{422}$ & $\chain{\rm m}{\bf 45}$ & $G_{421}$ & $\chain{\rm m}{\bf 45}$ & $G_{3211} $ & $\chain{\rm s}{\overline{\bf 126}}$ & & \cmark
\\[2mm]
\hline
\end{tabular}
\caption*{{\bf Table~\ref{tab:chains_2}} (Continued). }
\end{table}

The spontaneous symmetry breaking of $SO(10)$ and all intermediate symmetries are achieved by heavy Higgses, which acquire non-trivial vacuum expectation values (VEV) at the relevant scales. The $SO(10)$ Higgs multiplets used for this symmetry breaking are listed in the Tables~\ref{tab:chains_1} and \ref{tab:chains_2} below the arrows.
For example, a $\overline{\bf 126}$ Higgs of $SO(10)$ can be used to break $SO(10)$ or its subgroups to $G_{\rm SM}$, because this multiplet contains a trivial singlet of $G_{\rm SM}$ but not a trivial singlet of any larger symmetry group which contains $G_{\rm SM}$ as its subgroup. Once the $\overline{\bf 126}$ gains {a} VEV, it can break any larger group to $G_{\rm SM}$. 
A ${\bf 45}$ can be used to break any larger symmetry to $G_{3221}$ as it includes a component which is a parity-odd singlet of $G_{3221}$. 
 The ${\bf 45}$ includes another component which is a singlet of $G_{421}$ and could be used for the breaking to $G_{421}$. The ${\bf 54}$ includes a parity-even singlet of $G_{422}$, which is important for the breaking $SO(10) \to G_{422}^C$. Finally, the breaking $G_{422}^C \to G_{422}$ and $G_{422}^C \to G_{3221}^C$ can be achieved by including a ${\bf 210}$ which contains a parity-odd singlet of $G_{422}$ and another parity-even singlet of $G_{3221}$.

The formation of topological defects is ubiquitous in Grand Unified Theories. 
These defects are formed during the breaking of a larger gauge symmetry $G_I$ to a smaller one, $G_{I-1} \subset G_I$. 
The classification of topological defects is based on
the non-trivial homotopy group $\pi_k(G_I/G_{I-1})$. In particular, $\pi_2(G_I/G_{I-1}) \neq 0$ results in point-like monopoles, $\pi_1(G_I/G_{I-1}) \neq 0$ which leads to the formation of one-dimensional cosmic strings, and $\pi_0(G_I/G_{I-1}) \neq 0$ leads to the formation of two-dimensional domain walls.\footnote{The attributives ``point-like'', ``one-dimensional'' and ``two-dimensional'' refer only to cores of these topological defects, respectively.} In Tables~\ref{tab:chains_1} and \ref{tab:chains_2}, we list topological defects above arrows for each step of breaking in all chains, where ``m'', ``s'' and ``w'' denote monopoles, strings and domain walls, respectively. 

Monopoles and domain walls are cosmologically undesirable as their presence conflicts with our observed Universe. 
Monopoles have a heavy mass which is close to the energy scale of the symmetry breaking (ignoring the order-one gauge coupling of the GUT) and their number density is inversely proportional to the horizon volume.  Once generated, their mass and the total number does not change  as the Universe expands.  However, it is possible that they come to dominate the Universe's energy density, during radiation  domination, 
due to their heavy mass.
Their total mass is easy to dominate the Universe in the radiation domination era. 
Domain walls,\footnote{Here we focus on only the topologically stable domain walls. In non-GUT theories, explicit breaking terms may exist and domain walls become unstable and radiate GWs, see, e.g., \cite{Saikawa:2017hiv, Gelmini:2020bqg}.} as numerical simulations have shown (see e.g., a review \cite{Saikawa:2017hiv}), present a scaling behaviour after their production. The scaling solution results in the energy density of the walls  as $\rho_{\rm w} \propto \sigma H$, with $\sigma$ the tension of walls of mass dimension 3, leading to the energy density fraction $\Omega_{\rm w} \equiv \rho_{\rm w}/\rho_{\rm c} \propto G \sigma/ H$. Here, $\rho_{\rm c} = 3H^2/(8\pi G)$ is the critical energy density associating with the Hubble expansion rate $H$ and $G$ is the Newton constant. As the Universe expansion decelerates, the domain walls easily dominate the Universe. 
An era of inflation, which occurs during or after these defect's formation, can rid the Universe of them and solve this cosmological problem.
The cosmic strings network also obeys the scaling solution, leading to $\rho_{\rm s} \propto \mu H^2$. Then, $\Omega_{\rm s} \equiv \rho_{\rm s}/\rho_{\rm c} \propto G \mu$ remains as a small constant, which does not course  this cosmological problem.

The presence of topologically stable strings generated in the early Universe can be observable due to the gravitational radiation.\footnote{On the other hand, cosmic strings from $SO(10)$ GUT symmetry breaking may be unstable or metastable due to theoretical-field decay in consequence of GUT monopole nucleation \cite{Vilenkin:1982hm}, which eventually results in the disappearance of the entire network after a finite lifetime. In this kind of model, the large separation between the GUT scale and the energy scale that controls the string tension will then result in a long lifetime of the string network \cite{Dror:2019syi}. Monopole nucleation will occur and result in the breaking of long string segments on super-horizon scales, but this will have little phenomenological consequences because of the exponentially suppressed decay rate \cite{Buchmuller:2021mbb}. Stable network of Nambu–Goto string is then considered as an approximation that is justified by the hierarchy of scales in GUT symmetry breaking chain.} The picture is briefly described as follows: strings are generated after the symmetry breaking and a string network is formed during the cosmological expansion; strings collide and closed loops forms; the energy loss of the network via the emission of GWs simply corresponds to its way of maintaining the scaling regime during radiation or matter domination.

In the last columns of Tables~\ref{tab:chains_1} and \ref{tab:chains_2}, we indicate if, in principle, it is possible to observe cosmic strings via the measurement of GW background. In the final step of intermediate symmetry breaking, i.e., $G_1 \to G_{\rm SM}$, if gravitationally stable defects (indicated by either ``m'' or ``w'' or both) are generated, the inflation era has to be introduced during or after the last step of breaking. In this case, cosmic strings, if they are generated earlier, would be inflated away and thus, it is not observable. This case is marked as a ``\xmark\,'' in the last column. Otherwise, if the cosmic strings are generated alone in the last step of breaking, an inflationary stage can be introduced earlier than the formation of strings, and GW background from the string network may be observable. We mark this case as ``\cmark'' in the last column. Finally, we note that type (d) chains always produce unwanted defects in the final stage of spontaneous symmetry breaking. Consequently, a period of inflation is needed to eliminate these unwanted defects and would thus dilute away
any pre-existing cosmic string network associated GW signal. Therefore, we do not study type (d) chains. 
Note that in some special cases, the inflation and string formation may take place synchronously. Such cases may lead to diluted but observable strings \cite{Guedes:2018afo,Cui:2019kkd}. In our earlier work, this scenario was discussed \cite{King:2020hyd}; however, we will not include this possibility in our present discussion due to its added complexity but relegate it for future study. 

\section{The unification of gauge couplings}\label{sec:unificationRGE}

From our classification of $SO(10)$ breaking chains outlined in \secref{sec:classify}, we can solve the RGE for each breaking chain of type (c) to predict the associated scale of intermediate symmetry breaking and proton lifetime. The details of the RG running are provided in \secref{sec:RGR} and we
discuss the correlation of GUT and intermediate scales, which determines the scale of proton lifetime and gravitational waves, in \secref{sec:correlation}.
\subsection{RG running equations}\label{sec:RGR}

Any intermediate symmetry after $SO(10)$ breaking can be written as a product of a series of Lie groups $H_1 \times \cdots \times H_n$ or Lie groups combined with $Z_2^C$ as shown in Eq.~\eqref{eq:symmetries}. We denote the gauge coupling of the Lie group $H_i$ as $g_i$. The two-loop RG running equation for $g_i$ is given by
\begin{eqnarray}
\mu \frac{dg_i}{d\mu} = g_i \beta_i\,,
\end{eqnarray}
with $\beta_i$ determined by the particle content of the theory:
\begin{eqnarray}
\beta_i = \frac{g_i^2}{(4\pi)^2} \left\{ b_i + \sum_j b_{ij} \frac{g_j^2}{(4\pi)^2} \right\}\,. 
\end{eqnarray}
Throughout, we will ignored the contribution from Yukawa couplings to the 
RG running. In the case that $H_i$ and $H_j$ are non-Abelian groups, the coefficients of the beta function are given by
\begin{equation}
\begin{aligned}
b_i &=- \frac{11}{3} C_2(H_i) + \frac{2}{3} \sum_F T(F_i) + \frac{1}{3} \sum_S T(S_i) \,,\\
b_{ij} &=
- \frac{34}{3} [C_2(H_i)]^2 \delta_{ij} + \sum_F T(F_i) [2 C_2(F_j) + \frac{10}{3} C_2(H_i) \delta_{ij}] + \sum_S T(S_i) [4 C_2(S_j) + \frac{2}{3} C_2(H_i) \delta_{ij}]\,,
\end{aligned}
\end{equation}
where $F$ and $S$ represent chiral fermion and complex scalar multiplets respectively and $F_i$ and $S_i$ are their representations in the group $H_i$.
The quadratic Casimir of representation $R_i$ of the group $H_i$ is denoted as $C_2(R_i)$ for $R_i=F_i, S_i$. While the quadratic Casimir of the adjoint presentation of the group $H_i$ is directly denoted as $C_2(H_i)$. For this paper, it is important to note that

\begin{table}[t!]
\centering
\begin{tabular}{|c|cc|}
\hline\hline
Chains & $M_X$ [GeV] & $M_1$ [GeV] \\\hline
I1 & $5.660 \times 10^{15}$& $1.617 \times 10^{10}$ \\ 
I2 & $1.410 \times 10^{15}$ & $8.630 \times 10^{10}$ \\
I3 & $2.902 \times 10^{14}$ & $1.634 \times 10^{11}$ \\
I4 & $3.500 \times 10^{16}$ & $4.368\times 10^{9}$ \\ 
I5 & $2.722 \times 10^{14}$ & $1.143 \times 10^{13}$ \\
I6 & \multicolumn{2}{c|}{excluded} \\
\hline\hline
\end{tabular}
\caption{Predictions of the GUT scale, $M_X$, and the lowest intermediate scale, $M_1$, in breaking chains with a single intermediate scale. Chain I6 is excluded as no solution for the gauge unification exists. \label{tab:M_I}}
\end{table}

\begin{itemize}
\item For $SU(N)$, $C_2(SU(N))=N$ and the quadratic Casimir of the fundamental irreducible representation ${\bf N}$ of $SU(N)$ is given by $C_2({\bf N}) = (N^2-1)/2N$.
\item For $SO(10)$, $C_2(SO(10))=8$ and the quadratic Casimir of the fundamental irreducible representation ${\bf 10}$ of $SO(10)$ is given by $C_2({\bf 10}) = 9/2$. The spinor representation
of $SO(10)$ is ${\bf 16}$ and $C_2({\bf 16}) = 45/4$.
\end{itemize}
 $T(R_i)$ is the Dynkin index of representation $R_i$ of group $H_i$. In particular for $SU(N)$, $T(R_i) = C_2(R_i)d(R_i)/(N^2-1)$ where $d(R_i)$ is the dimension of $R_i$.
If a single $H_j$ is a $U(1)$ symmetry, the coefficient $b_{ij}$ is obtained by replacing $C_2(R_j)$ and $T(R_j)$ with the charge square $[Q_j(R)]^2$ of the field multiplet $R$ in $U(1)_j$.
In the case that $H_i$ is an Abelian $U(1)$ symmetry, its beta function is also simply modified by replacing both $C_2(R_i)$ and $T(R_i)$ with the charge square $[Q_i(R)]^2$. For the Abelian symmetry, $C_2(U(1)) = 0$.
By denoting $\alpha_i =g_i^2/(4\pi)$, the RG running equation can be rewritten as
$\mu d\alpha_i/d\mu = \widetilde{\beta}_i (\alpha_i)$ where $\widetilde{\beta}_i$ is given by
\begin{eqnarray}
\widetilde{\beta}_i = \frac{1}{2\pi} \alpha_i^2 ( b_i + \frac{1}{4\pi} \sum_{j} b_{ij} \alpha_j ) \,.
\end{eqnarray}
The coefficients of the beta functions, $b_i$ and $b_{ij}$, at the one- and two-loop level, respectively, are provided in Table~\ref{tab:beta_coefficients} of Appendix~\ref{app:beta_coefficients}. Note that the values of $b_i$ and $b_{ij}$ depend on the degrees of freedom introduced for gauge, matter and Higgs fields. As discussed, the gauge fields are directly determined by the gauge symmetry pattern, and we assume the matter fields to include all the SM fermions and right-handed neutrinos. The largest uncertainty in the particle content comes from the Higgs fields. 
However, we follow the criterion mentioned in the introduction and derive the minimal required Higgs multiplet content needed for each breaking chain and each intermediate symmetry. We list the Higgs representation in Appendix~\ref{app:beta_coefficients} explicitly.

The RGE at two-loop can be solved analytically \cite{Bertolini:2009qj}:
\begin{eqnarray}
\alpha_i(\mu)^{-1} = \alpha_i(\mu_0)^{-1} - \frac{b_i}{2\pi}\log\frac{\mu}{\mu_0} + \sum_j \frac{b_{ij}}{4\pi b_i} \log\left(1- b_j \alpha_j(\mu_0) \log \frac{\mu}{\mu_0}\right) \,,
\end{eqnarray}
and this solution is valid for $b_j \alpha_j(\mu_0) \log(\mu/\mu_0) <1$. Given any breaking chain $SO(10) \to \cdots G_I \to G_{I-1} \to \cdots G_{\rm SM}$, gauge couplings of the 
 symmetry before the breaking (i.e., $G_I$) and the residual symmetry (i.e., $G_{I-1}$) after the breaking satisfies the matching conditions at the breaking scale $\mu=M_I$. In particular, for a simple Lie group $H_i \subset G_I$ broken to its subgroup $H_j$ which is also simple Lie group and $H_j \subset G_{I-1}$, the one-loop matching condition at $\mu=M_I$ is given by
\begin{equation}\label{eq:match1}
H_{i} \to H_j\,, \quad
\frac{1}{\alpha_{H_i}(M_I)} - \frac{C_2(H_i)}{12\pi} = \frac{1}{\alpha_{H_j}(M_I)} -\frac{C_2(H_j)}{12\pi} \,.
\end{equation}
In the $SO(10)$ GUT symmetry breaking chains, we also encounter the breaking of $U(1) \times U(1) \to U(1)$ at the lowest intermediate scale $M_1$ and the matching condition depends on the $U(1)$ charges:
\begin{equation}\label{eq:match2}
U(1)_{R} \times U(1)_{X} \to U(1)_Y \,, \quad
\frac{3}{5\alpha_{1R}(M_1)} + \frac{2}{5\alpha_{1,X}(M_1)} = \frac{1}{\alpha_{1Y}(M_1)} \,. 
\end{equation}

\subsection{Correlation between the GUT scale and intermediate scales}\label{sec:correlation}

Using the matching conditions of Eqs.~\eqref{eq:match1} and \eqref{eq:match2}, all gauge couplings of the subgroups unify into a single gauge coupling, $g_X$, of $SO(10)$ at the GUT scale. i.e., all $\alpha_i$ are united into $\alpha_X \equiv g_X^2/{4\pi}$. This unification 
restricts both the GUT and intermediate scales for each breaking chain. 
We denote the mass of heavy gauge bosons associated with $SO(10)$ breaking as $M_X$. While the 
gauge boson masses $M_1,M_2,\cdots$ from the breaking of intermediate symmetries $G_1, G_2, \cdots$ are referred to as the scales of intermediate symmetries.
In this subsection, we will explore the correlation between the GUT scale, $M_X$, and intermediate scales, $M_{3}, M_{2}, M_{1}$. In particular, we focus on the correlation of the GUT and lowest intermediate scale, $M_1$.

We numerically solve the two-loop RG equations from the electroweak to the GUT scale. For example, for the breaking chain $SO(10) \to \cdots \to G_2 \to G_1 \to G_{\rm SM}$, the RG running procedure is performed in reverse: $G_{\rm SM} \to G_1 \to G_2 \to \cdots \to SO(10)$. 
We begin the evaluation from the $M_Z$ pole, where the three gauge couplings $\alpha_3$, $\alpha_2$ and $\alpha_1$ are 
\begin{eqnarray}
\alpha_3 = 0.1184\,,\quad
\alpha_2 = 0.033819\,, \quad
\alpha_1 = 0.010168 \,,
\end{eqnarray} 
at their best fit points \cite{Xing:2011aa}. These couplings are evolved using the RGE of the SM to scale $M_1$, where $G_1$ is recovered. Aided by the matching conditions for the gauge couplings $\alpha_3$, $\alpha_2$ and $\alpha_1$ of the SM and gauge couplings of the gauge symmetry, $G_1$, we obtain the values of couplings in the intermediate symmetry group. We then run the gauge couplings of $G_1$ from $M_1$ to the scale $M_2$, where the larger group, $G_2$, is recovered and the gauge couplings of $G_2$ are obtained via matching conditions at scale $M_2$. Repeating this same procedure, we run all couplings to the GUT scale to unify to a single value.
 In each breaking chain, the values of the $\beta$ coefficients used for the running between two neighbouring scales depend on particles content introduced in the theory. 
 Our treatment of the RG running is economical: we consider a non-supersymmetric theory and ignore the contribution of threshold effects of additional heavy particles. 
 
For breaking chains with a single intermediate scale, $M_1$, the GUT scale, $M_X$, can be uniquely determined by unifying the three SM gauge couplings. The predictions of all type (c) chains with a single intermediate scale are shown in Table~\ref{tab:M_I}. For chain I6, no solution provides gauge unification, and we will not consider this chain further.
\begin{figure}[t!]
\centering
\includegraphics[width=.92\textwidth]{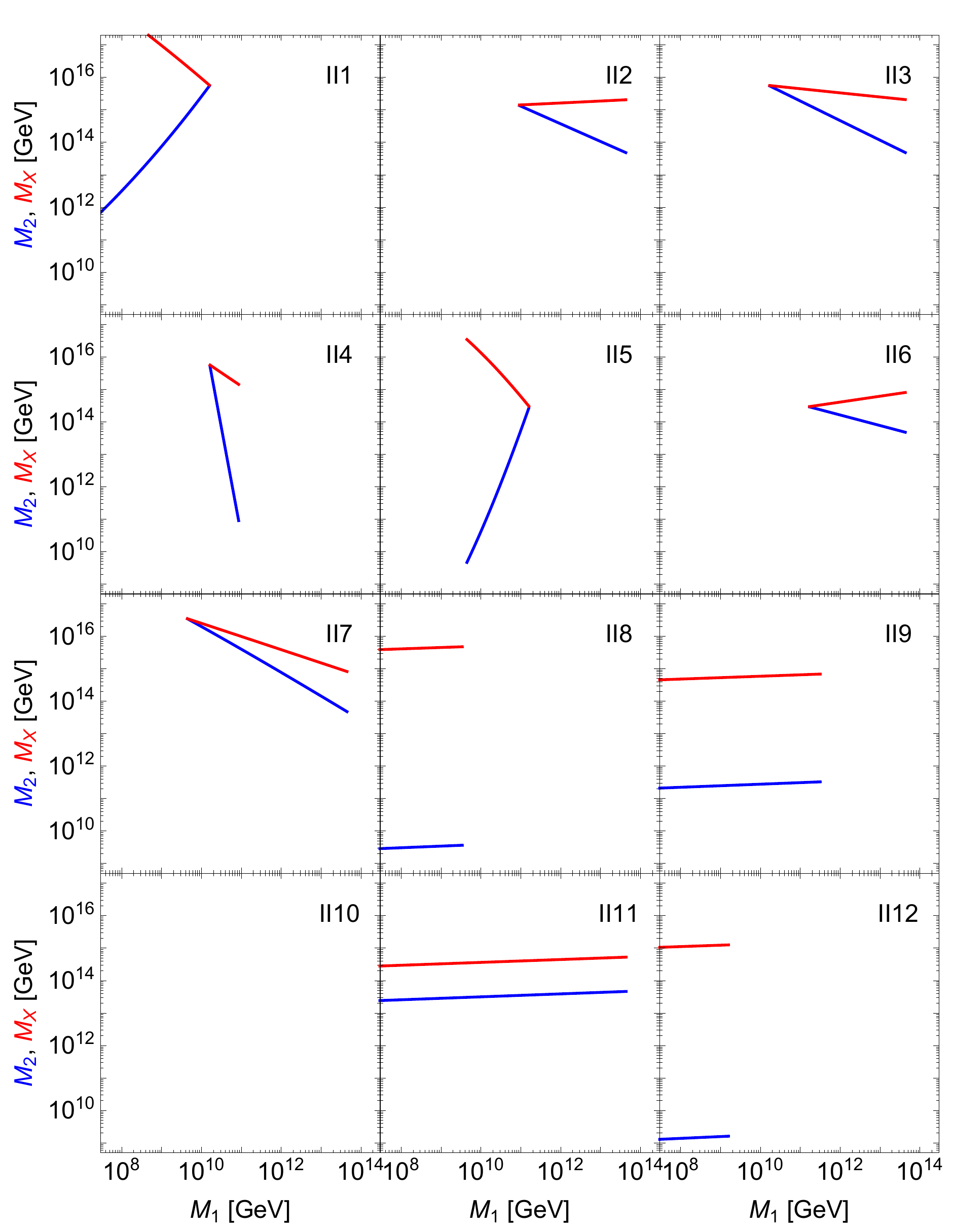}
\caption{Predictions of GUT scale, $M_X$, and intermediate scale, $M_1$, and $M_2$, for breaking chains with two intermediate scales: $SO(10) \to G_2 \to G_1 \to G_{\rm SM}$. Chain II10 is excluded as there exists no solution for gauge unification. We set the lower and upper bound on the mass scales to be $M_{\rm low} = 5 \times 10^7$~GeV and $M_{\rm pl}$ in the scan respectively. The energy scale ordering $M_{\rm low} \leqslant M_1 \leqslant M_2 \leqslant M_X \leqslant M_{\rm pl}$ is required in all cases.}\label{fig:RG_II}
\end{figure}

\begin{figure}[t!]
\centering
\includegraphics[width=.92\textwidth]{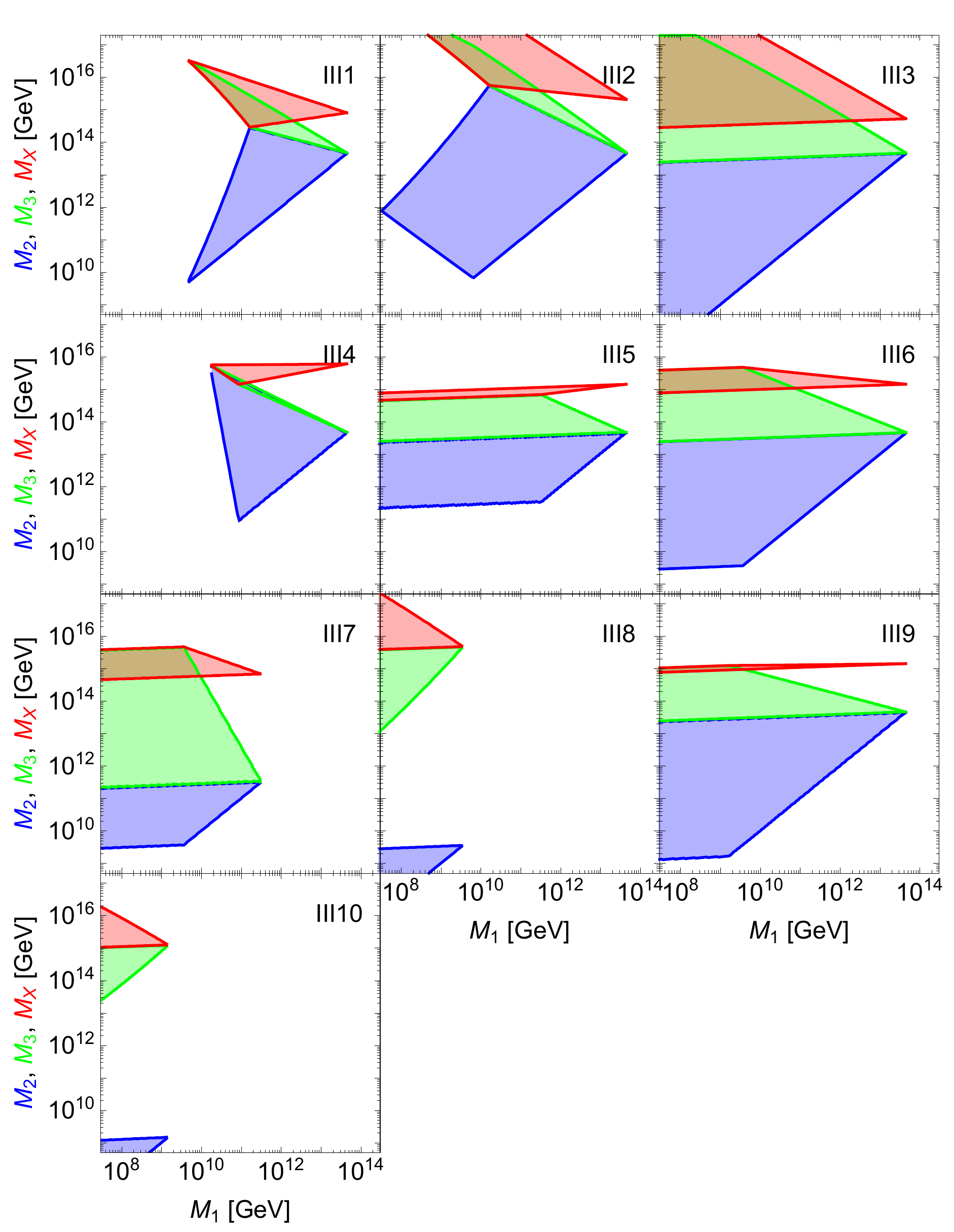}
\caption{Prediction of GUT scale, $M_X$, and intermediate scales $M_1$, $M_2$ and $M_3$, in type (c) breaking chains with three intermediate scales $SO(10) \xrightarrow{M_X} G_3 \xrightarrow{M_3} G_2 \xrightarrow{M_2} G_1 \xrightarrow{M_1} G_{\rm SM}$. The energy scale ordering $M_1 \leqslant M_2 \leqslant M_3 \leqslant M_X$ is required in all cases. We set the lower and bounds of all energy scales as $M_{\rm low} = 5 \times 10^7$~GeV and $M_{\rm pl}$, respectively, in the scan.}\label{fig:RG_III}
\end{figure}

For breaking chains with more than one intermediate scales, restriction to the scales are further relaxed as more free parameters are introduced. We scan over all intermediate scales and the GUT scale in a fixed interval $[M_{\rm low},~ M_{\rm pl}]$, where the lower bound is $M_{\rm low} = 5 \times 10^7$~GeV and the upper bound is the Planck scale, $M_{\rm pl} =1.22 \times 10^{19}$~GeV. More specifically, the scale ordering $M_{\rm low} \leqslant M_1 \leqslant M_2 \cdots M_X \leqslant M_{\rm pl}$ is always satisfied in our scans. There may exist a breaking chain with intermediate scales lower than $M_{\rm low}$. However, such a case will not be included in our scan. 

There is a single free parameter for breaking chains with two intermediate scales, which we choose to be the lowest intermediate scale, $M_1$. Given a fixed value of $M_1$, the condition of gauge unification determines the values of $M_X$ and $M_2$. By varying $M_1$ from $M_{\rm low}$ to $M_{\rm pl}$, we obtain the correlation between $M_1$ and $M_X$, $M_2$. In Fig.~\ref{fig:RG_II}, we show the correlation between $M_X$ and $M_2$ as functions of $M_1$ in the red and blue curves respectively. Two features merit discussion:
\begin{itemize}
\item In chains II2-7, the two curves intersect at a single point. This occurs when the two intermediate scales reduce to a single intermediate scale. For example, in chain II2: $SO(10) \to G_{422}^C \to G_{3221}^C \to G_{\rm SM}$, the intersecting point refers to chain I2: $SO(10) \to G_{3221}^C \to G_{\rm SM}$. We note that the predictions of $M_1$ and $M_X$ at this intersecting point match with those in chain I2 as listed in Table~\ref{tab:M_I}.
\item In all chains, the value of $M_2$ at one of the endpoints of the blue curve equals the value of $M_1$ at that point. At this point, the two intermediate scale chains reduce to single intermediate scale chains. For example, in chain II2, the left endpoint refers to chain I5, i.e., $SO(10) \to G_{422}^C \to G_{\rm SM}$. 

\end{itemize}

\begin{figure}[t!]
\centering
\includegraphics[width=.92\textwidth]{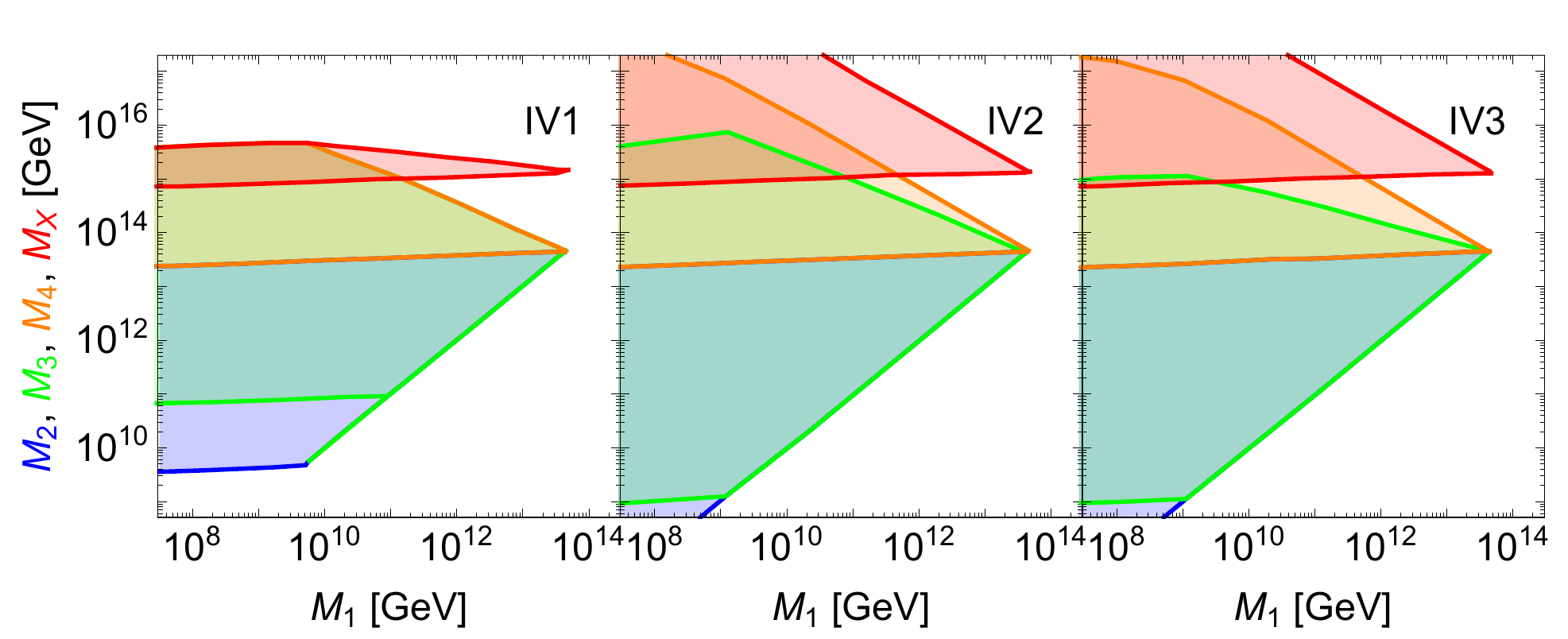}
\caption{Prediction of GUT scale, $M_X$, and intermediate scales, $M_1$, $M_2$, $M_3$ and $M_4$, in type (c) breaking chains with four intermediate scales: $SO(10) \xrightarrow{M_X} G_4 \xrightarrow{M_4} G_3 \xrightarrow{M_3} G_2 \xrightarrow{M_2} G_1 \xrightarrow{M_1} G_{\rm SM}$. The energy scale ordering $M_1 \leqslant M_2 \leqslant M_3 \leqslant M_4 \leqslant M_X$ is required in all cases. We set the lower and bounds of all energy scales as $M_{\rm low} = 5 \times 10^7$~GeV and $M_{\rm pl}$, respectively, in the scan.
The panels left to right are for the chains cIV1-3, respectively.}\label{fig:RG_IV}
\end{figure}

The above procedure provides a simple consistency check on the breaking chains with more than one intermediate scale.
In the case of three intermediate scales, there are even more free parameters. As such, fixing $M_1$ cannot determine the remaining scales. Instead, after $M_1$ is fixed, we can vary $M_X$ from $M_1$ to $M_{\rm pl}$ to determine $M_2$ and $M_3$, with the hierarchy $M_{\rm low} \leqslant M_1 \leqslant M_2 \leqslant M_3 \leqslant M_X \leqslant M_{\rm pl}$ required. From this we obtain a range for $M_X$, as well as ranges for $M_2$ and $M_3$, for a fixed $M_1$. By varying $M_1$ from $M_{\rm low}$ to $M_{\rm pl}$, we obtain a range of $M_X$ values, as well as ranges of $M_2$ and $M_3$, allowed by gauge unification. The results are shown in Fig.~\ref{fig:RG_III}. Some borders of these regions refer to the limiting case with only two intermediate scales. In particular, the border between the blue and green regions in chain III1 refers to $M_2$ in chain II6, and the red border just above it refers to $M_X$ in the same breaking chain. However, it is worth noting that this property does not always hold. Compared with type-II chains, the involvement of a further intermediate scale requires more Higgs fields for the new intermediate symmetry breaking. These Higgses may not be present in the particle content in relevant type-II chains and will provide extra contributions to the radiative corrections and modify the RG behaviour. Therefore, these type-III chains may not always be reduced to the relevant type-II chains when intermediate scales become degenerate. The results of breaking chains with four intermediate scales are given in Fig.~\ref{fig:RG_IV}. The inclusion of an additional intermediate scale enlarges the allowed parameter space of these scales even further.


\section{Proton decay and its constraints on GUT intermediate scales}\label{sec:PDresults}

As discussed before, for a given breaking chain, the solutions of the RGEs and the condition of gauge unification restricts the GUT scale and correlates it with the intermediate scales. As the proton decay rate is proportional to the GUT scale, we can use limits on this observable to constrain the GUT and intermediate scales, including the lowest intermediate scale $M_1$. In this Section, we discuss
 how we calculate the proton lifetime given a prediction of the GUT scale.

 GUT scale physics can induce proton decay via four dimension-six operators present at the electroweak (EW) scale:
\begin{equation} \label{eq:D6}
\begin{aligned}
=\epsilon^{ijk} \epsilon_{\alpha\beta} \Big(&\frac{1}{\Lambda_1^2}
(\overline{u_R^{jc}} \gamma^\mu Q^k_\alpha)(\overline{d_R^{ic}} \gamma_\mu L_\beta) +
\frac{1}{\Lambda_1^2}
(\overline{u_R^{jc}} \gamma^\mu Q^k_\alpha)(\overline{e_R^{c}} \gamma_\mu Q_\beta^i)\\
+ &\frac{1}{\Lambda_2^2}
(\overline{d_R^{jc}} \gamma^\mu Q^k_\alpha)(\overline{u_R^{ic}} \gamma_\mu L_\beta) +
\frac{1}{\Lambda_2^2}
(\overline{d_R^{jc}} \gamma^\mu Q^k_\alpha)(\overline{\nu_R^{c}} \gamma_\mu Q^i_\beta)
+{\rm h.c.} \Big) \,,
\end{aligned}
\end{equation}
where $i,j,k$ ($\alpha,\beta$) denotes colour (flavour) indices and $\Lambda_{1}, \Lambda_{2}$ are the UV completion scales of the GUT symmetry \cite{Weinberg:1979sa,Wilczek:1979hc,Weinberg:1980bf,Weinberg:1981wj,Sakai:1981pk}. 
For type (c) breaking chains: $\Lambda_1 = \Lambda_2 \simeq \left(g_X M_X\right)/2$
and these operators induce the proton decay into a meson and a lepton (anti-lepton) and
the golden channel is $p \to \pi^0 e^+$ which has the decay width:
\begin{equation}
\begin{aligned}
\Gamma (p \to \pi^0 + e^+) =
\frac{m_p}{32 \pi} \left( 1- \frac{m_{\pi^0}^2}{m_p^2} \right)^2 A_L^2 \times
\Big[
&A_{SL} \Lambda_1^{-2} (1+ |V_{ud}|^2) \, |\langle \pi^0 \left| (ud)_R u_L | p \rangle \right|^2
\Big.\\
+
&\Big.A_{SR} (\Lambda_1^{-2}+ |V_{ud}|^2 \Lambda_2^{-2}) \, \left|\langle \pi^0 | (ud)_L u_L | p \rangle \right|^2
\Big]\,,
\end{aligned}
\end{equation}
where $A_L$, $A_{SL}$ and $A_{SR}$ enhancement factors induced by the long and short range effects on proton decay respectively. The hadronic matrix element relevant for our decay mode is $\langle \pi^0 | (ud)_{L,R} u_L | p \rangle$, and this has been obtained from a QCD lattice simulation \cite{Aoki:2017puj}.
 The long range effect account for the renormalisation enhancement from the proton decay ($m_p\sim 1$~GeV) to the EW scale (set to be the $Z$ mass at scale $M_Z$) which is $A_L =1.247$ calculated at two-loop level \cite{Nihei:1994tx,Ellis:2020qad}.
 
\begin{table}[t!]
\centering
\begin{tabular}{|l|c|c|}
\hline\hline
\multirow{2}{2cm}{Intermediate symmetry}&\multicolumn{2}{c|}{Anomalous dimensions} \\\cline{2-3}
& $\{ \gamma_{iL} \}$ & $\{ \gamma_{iR} \}$ \\\hline
&&\\[-4mm]
$G_{321}$ & $\left\{2,\frac{9}{4},\frac{23}{20}\right\}$ & $\left\{2,\frac{9}{4},\frac{11}{20}\right\}$ \\[1mm]\hline
&&\\[-4mm]
$G_{3211}$ & $\left\{2,\frac{9}{4},\frac{3}{4},\frac{1}{4}\right\}$ & $\left\{2,\frac{9}{4},\frac{3}{4},\frac{1}{4}\right\}$ \\[1mm]\hline
&&\\[-4mm]
$G_{3221}^{(C)}$ & $\left\{2,\frac{9}{4},\frac{9}{4},\frac{1}{4}\right\}$ & $\left\{2,\frac{9}{4},\frac{9}{4},\frac{1}{4}\right\}$ \\[1mm]\hline
&&\\[-4mm]
$G_{421}$ & $\left\{\frac{15}{4},\frac{9}{4},\frac{3}{4}\right\}$ & $\left\{\frac{15}{4},\frac{9}{4},\frac{3}{4}\right\}$ \\[1mm]\hline
&&\\[-4mm]
$G_{422}^{(C)}$ & $\left\{\frac{15}{4},\frac{9}{4},\frac{9}{4}\right\}$ & $\left\{\frac{15}{4},\frac{9}{4},\frac{9}{4}\right\}$ \\[1mm]\hline\hline
\end{tabular}
\caption{Anomalous dimensions for intermediate symmetries of $SO(10)$ breaking to $G_{321}$. }\label{tab:anomalous}
\end{table}

\begin{figure}[t!]
\centering
\includegraphics[width=.78\textwidth]{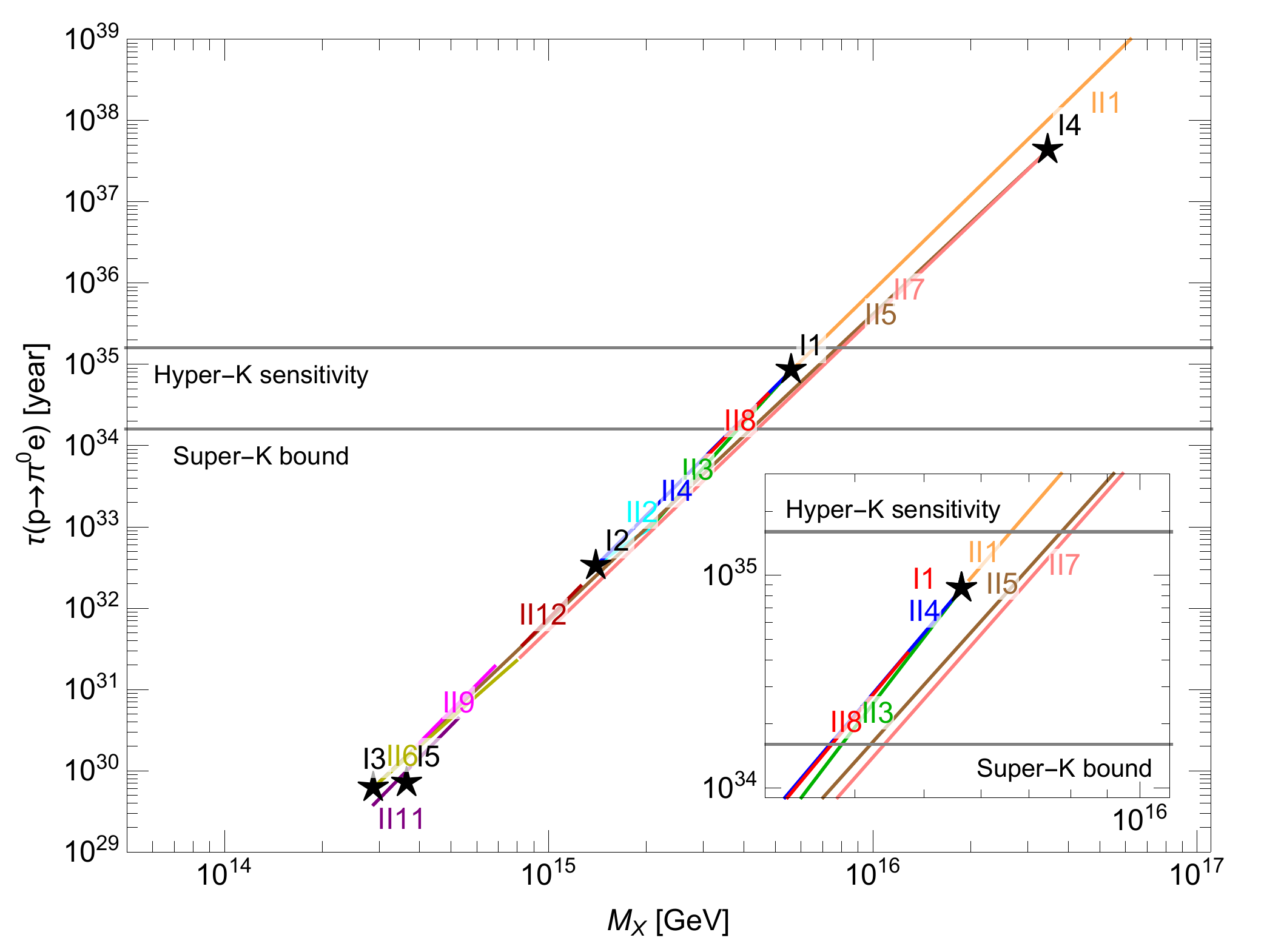}
\includegraphics[width=.75\textwidth]{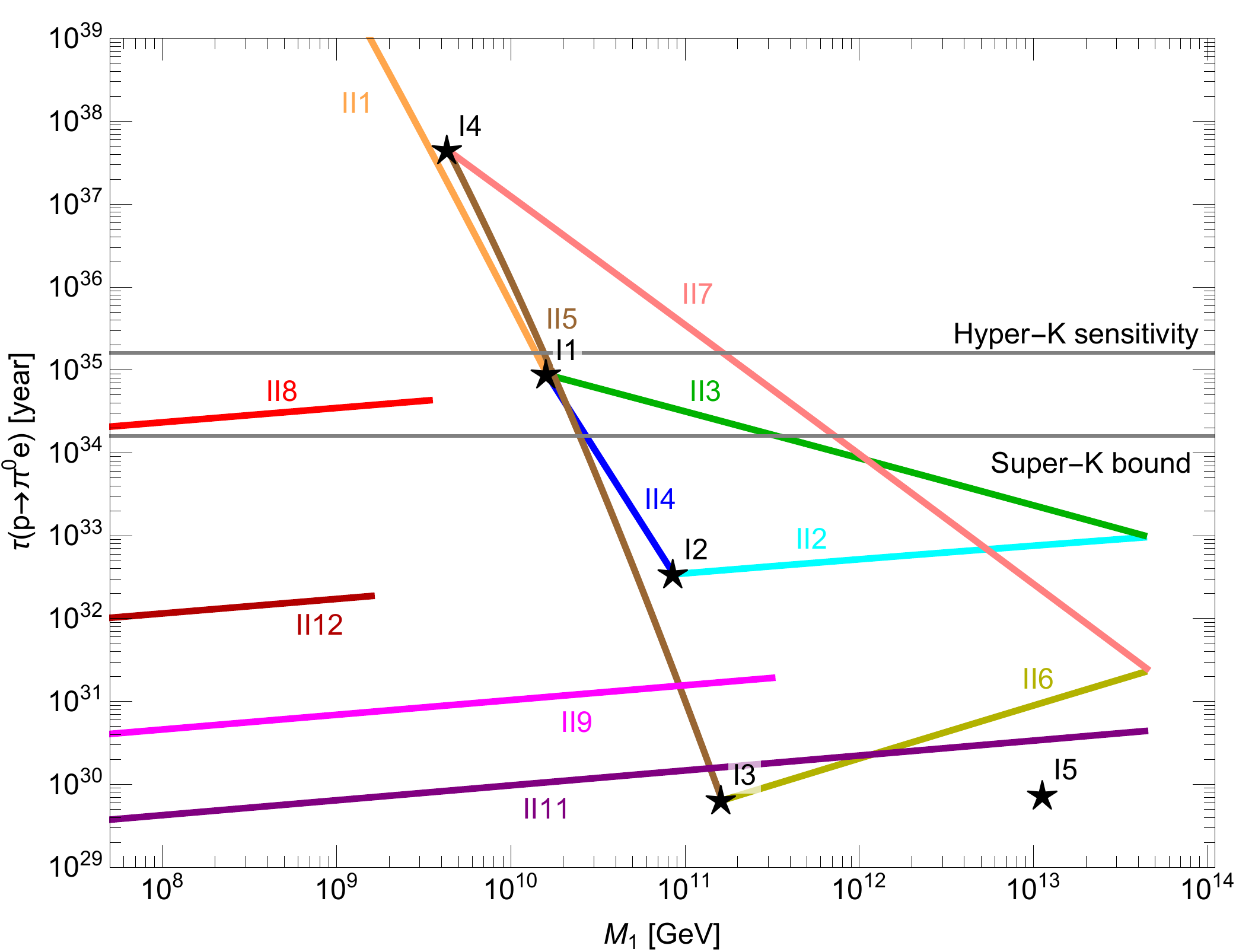}
\caption{Predictions of the proton decay lifetime as a function of the GUT scale, $M_X$, and intermediate scale, $M_1$, in breaking chains 
cI1-5 and cII1-12. The colour for each chain is specified in both the upper and low panel. Labels for each chains in the upper panel are not shown as regions for these chains overlap significantly. }\label{fig:pd_cI_II}
\end{figure}

\begin{figure}[t!]
\centering
\includegraphics[width=.78\textwidth]{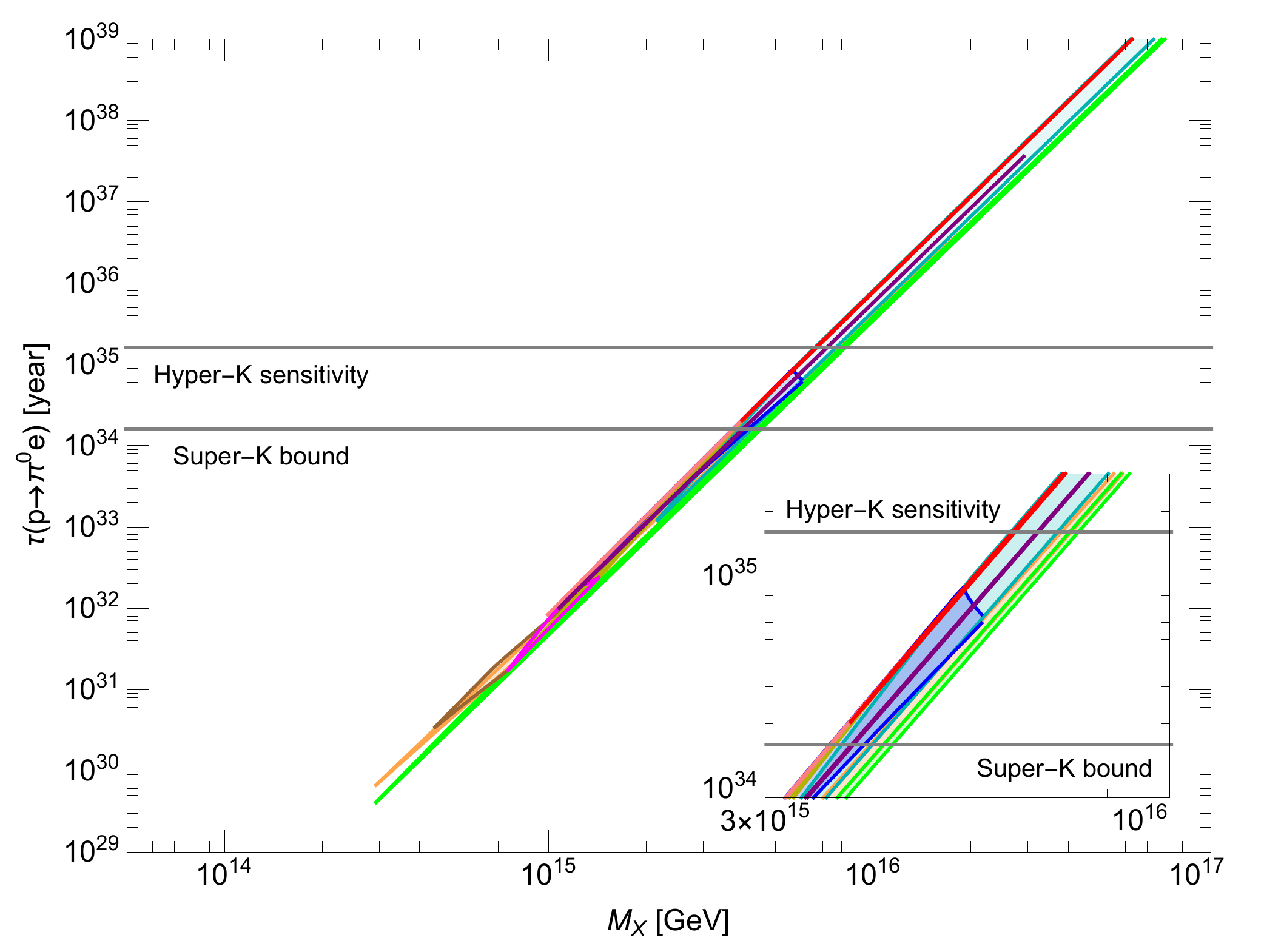}
\includegraphics[width=.75\textwidth]{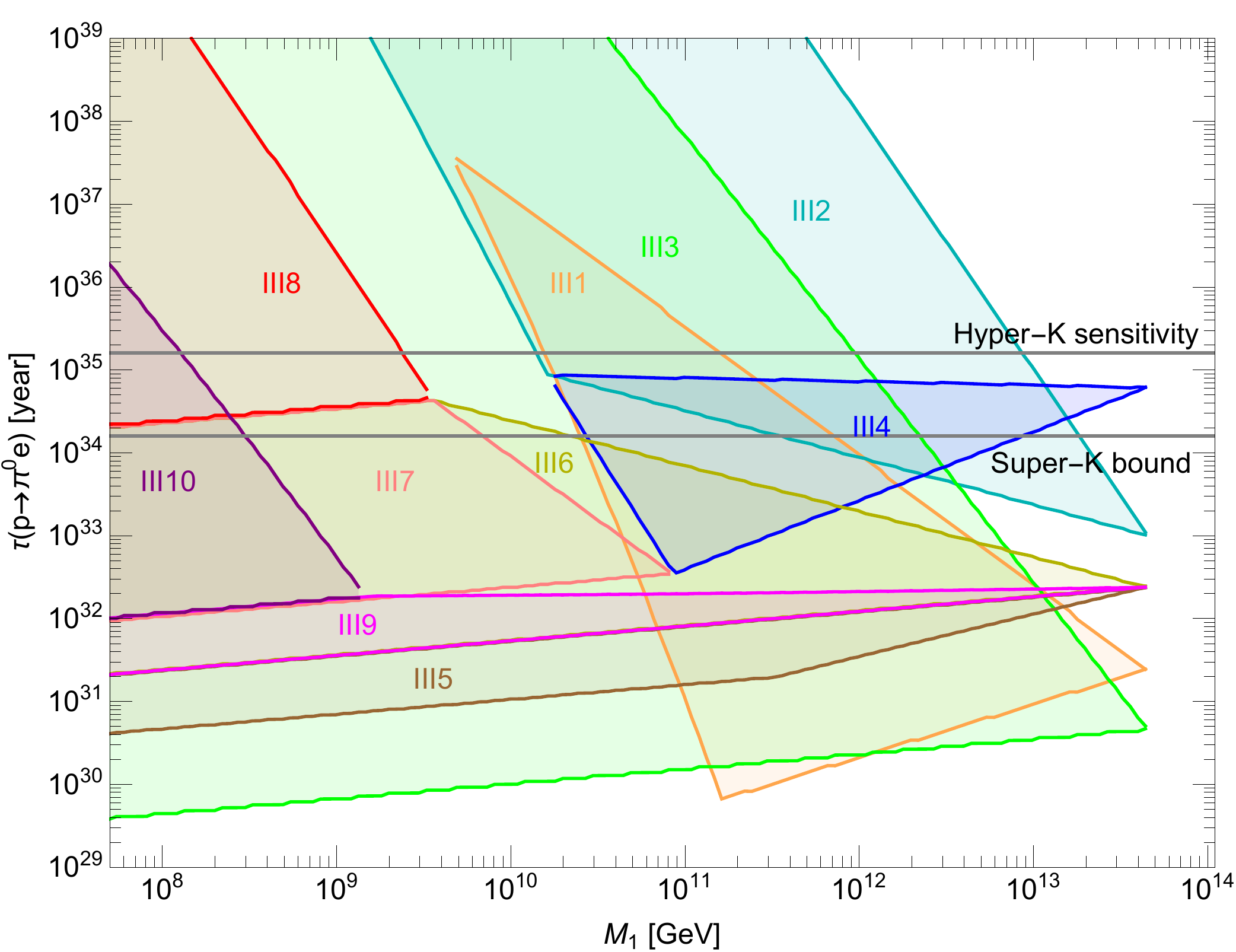}
\caption{Predictions of the proton decay lifetime as a function of the GUT scale, $M_X$, and intermediate scale, $M_1$, in breaking chains cIII1-10. The colour for each chain is specified in both the upper and low panel. Labels for each chains in the upper panel are not shown as regions for these chains overlap significantly. }\label{fig:pd_cIII}
\end{figure}

\begin{figure}[t!]
\centering
\includegraphics[width=.49\textwidth]{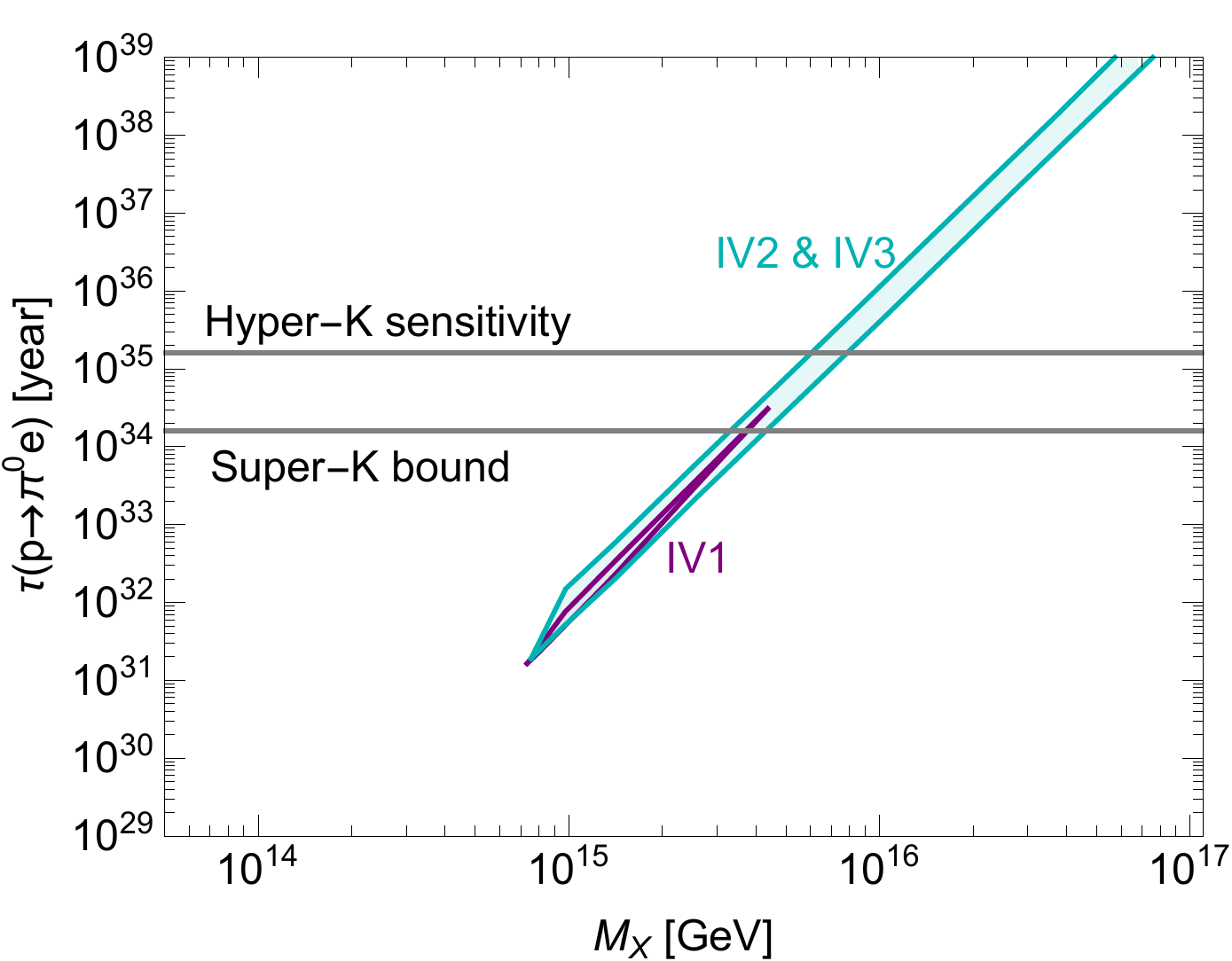}
\includegraphics[width=.49\textwidth]{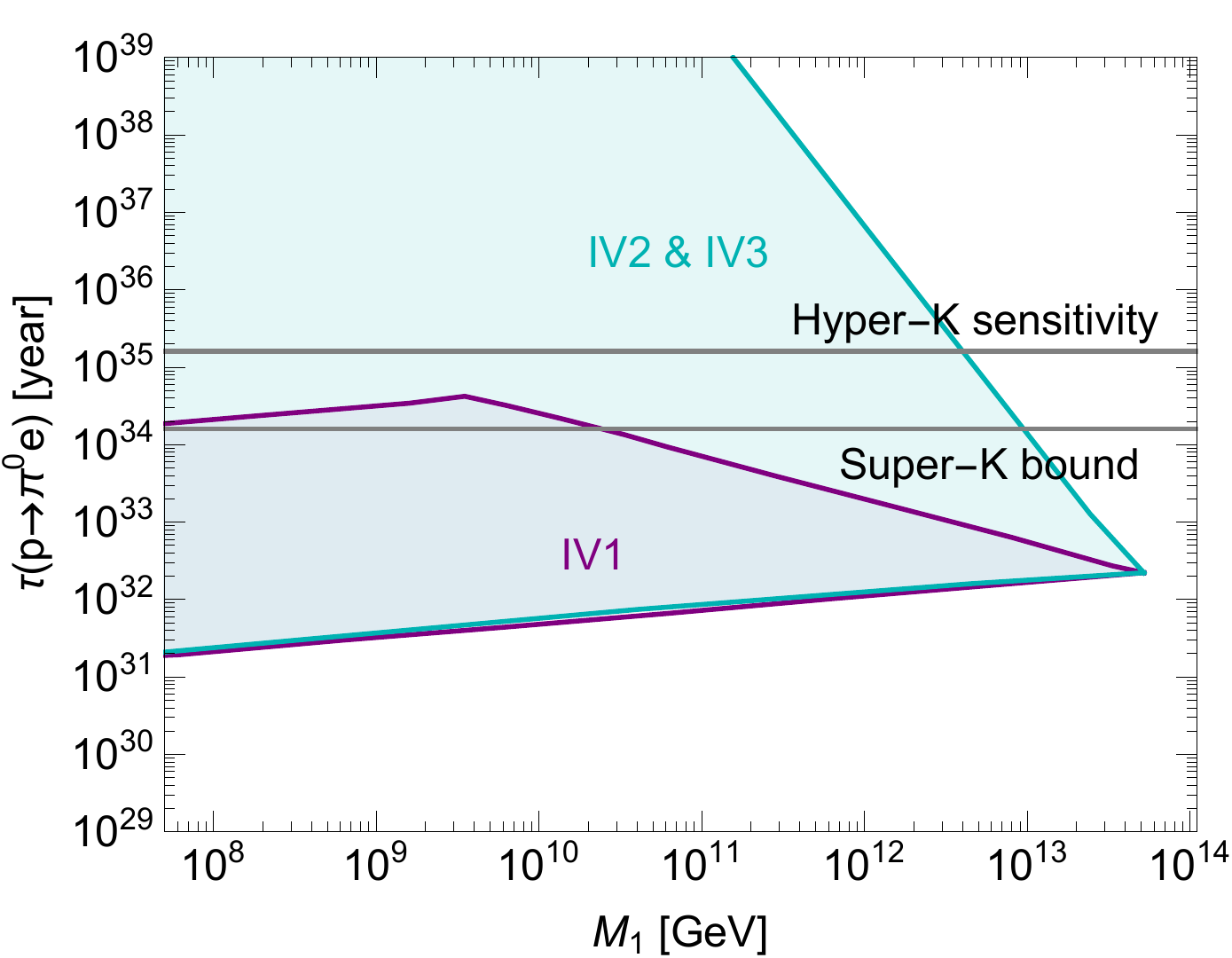}
\caption{Predictions of the proton decay lifetime as a function of the GUT scale, $M_X$, and intermediate scale, $M_1$, in breaking chains 
IV1-3. The colour for each chain is specified in both the upper and low panel. }\label{fig:pd_cIV}
\end{figure}
 
The factors associated with the short-range effects, $A_{SL}$ and $A_{SR}$, are obtained from the RG equation running from $M_Z$ to the scale $M_X$, which is the mass of the integrated baryon number violating mediators. Thus, these factors are non-trivially dependent upon the breaking chain. 
The short-range factors are parametrised as 
\begin{eqnarray}
A_{SL(R)} = \prod_A^{M_Z \leqslant M_A \leqslant M_X} \prod_i \left[ \frac{\alpha_i (M_{A+1})}{\alpha_i(M_A)} \right]^{\frac{\gamma_{iL(R)}}{b_i}}\,,
\end{eqnarray}
where $\gamma_i$ and $b_i$ denote the anomalous dimension and one-loop $\beta$ coefficient and those values at given intermediate scales are given in Tables~\ref{tab:anomalous} and \ref{tab:beta_coefficients} respectively. These short range effects can be numerically obtained using the RG running discussed in the previous section. 
We note that $M_X$ is correlated with $\Lambda_1$ or $\Lambda_2$ via $M_X \sim g_X \min \{\Lambda_1, \Lambda_2\}$.
From the discussion in the previous section, we are now able to predict the proton lifetime $\tau(p \to \pi^0 + e^+) = 1/\Gamma(p \to \pi^0 + e^+)$ for a given breaking chain.
The predictions of the proton lifetime for all breaking chains with one or two intermediate scales of type (c) are provided in Fig.~\ref{fig:pd_cI_II}. In the upper panel, we compare the proton lifetime with the GUT scale, $M_X$. Clearly, there is a power law correlation between $\tau$ and $M_X$,
\begin{eqnarray} \label{eq:power_law}
\tau \simeq 6.9 \times 10^{35} {\rm years} \times \left(\frac{M_X}{10^{16} {\rm GeV}}\right)^4 \,,
\end{eqnarray} 
which is approximately satisfied regardless of the breaking chain. A null result observed by Super-K provides a lower limit on the proton lifetime $\tau > 1.6 \times 10^{34}$ years. We can convert this limit to a lower bound on the GUT scale, $M_X \gtrsim 4 \times 10^{15}$~GeV. Moreover, as the gauge unification correlates $M_X$ with the lower intermediate scales, we can transform the $\tau$-$M_X$ correlation to a 
$\tau$-$M_1$ correlation which is shown in the lower panel of Fig.~\ref{fig:pd_cI_II}.

We find that the correlations between $\tau$ and $M_1$ can vary significantly for each chain.
The breaking chains which are allowed by the Super-K constraint are  
I1,4, and II1,3,4,5,7,8. Importantly, these chains will be tested by the future Hyper-K experiment. An observation of the $\pi^0 e^+$ signal in Hyper-K, depending on the value of the measured lifetime, will exclude some of these chains. However, if proton decay is not observed during Hyper-K's 10-year exposure time, chains I1, II3,4 and 8 will be excluded.

For breaking chains with three intermediate scales, the predictions of the proton lifetime as a function of $M_X$ and $M_1$ are shown in Fig.~\ref{fig:pd_cIII}. The correlation between $M_{2}$ and $\tau$ is a region rather than a line. 
In all these cases, the power-law $\tau \sim M_X^4$ remains a good approximation. We find that some areas of the parameter space of chains III1-4,6-8,10 are allowed, and Hyper-K will exclude III4,6,7 if proton decay is not observed.
Finally, the results for breaking chains with four intermediate scales are shown in Fig.~\ref{fig:pd_cIV}. All three chains (IV1,2,3) are still allowed by the Super-K constraint. IV1 would be excluded by the possible non-observation of proton decay by Hyper-K, but Hyper-K cannot entirely exclude IV2,3 due to their large parameter space given the additional intermediate scale. Regions for chains IV2,3 coincide, as shown in the plot. They differ only when a very long proton lifetime is predicted, which is out of the range shown in the plot. 

We finally summarise bounds on the GUT scale $M_X$ and predicted ranges of the lowest intermediate scale $M_1$ from Super-K and the future Hyper-K experiments in Table~\ref{tab:pd_scales}. Regions of $M_X$ and $M_1$ from Hyper-K are obtained by assuming a null result of the proton decay after a 10-year exposure. 

\section{Gravitational waves generated from cosmic string networks}\label{sec:GW1} 


\begin{table}[t!]
\centering
\resizebox{\textwidth}{!}{
\begin{tabular}{|c|cc|cc|c|}
\hline\hline
\multirow{2}{*}{Chain} & \multicolumn{2}{c|}{Super-K} & \multicolumn{2}{c|}{Hyper-K in the future} & GWs via \\
& bound\,on\,$M_X$\,[GeV] & predicted\,$M_1$\,[GeV] & 
bound\,on\,$M_X$\,[GeV] & predicted\,$M_1$\,[GeV] & strings \\\hline\hline
I1 & $M_X \simeq 5.6 \!\cdot\! 10^{15}$ & $M_1 \simeq 1.6 \!\cdot\! 10^{10}$ & \multicolumn{2}{c|}{will be excluded} & \cmark \\
I4 & $M_X \simeq 3.5 \!\cdot\! 10^{16}$ & $M_1 \simeq 4.4 \!\cdot\! 10^{9}$ & $M_X \simeq 3.5 \!\cdot\! 10^{16}$ & $M_1 \simeq 4.4 \!\cdot\! 10^{9}$ & \xmark \\\hline
II1: & $M_X \gtrsim 5.8 \!\cdot\! 10^{15}$ & $M_1 \lesssim 1.6\!\cdot\! 10^{10}$ & $M_X \gtrsim 6.5 \!\cdot\! 10^{15}$ & $M_1 \lesssim 1.4 \!\cdot\! 10^{10}$ & \cmark\\
II3: & $3.7\!\cdot\! 10^{15} \text{--}\, 3.8\!\cdot\! 10^{15}$ & $1.6\!\cdot\! 10^{10} \text{--}\, 3.3\!\cdot\! 10^{11}$ & \multicolumn{2}{c|}{will be excluded} & \cmark\\
II4: & $3.7\!\cdot\! 10^{15} \text{--}\, 5.6\!\cdot\!10^{15}$ & $1.6\!\cdot\! 10^{10} \text{--}\, 2.7\!\cdot\! 10^{10}$ & \multicolumn{2}{c|}{will be excluded} & \cmark\\
II5: & $4.2\!\cdot\! 10^{15} \text{--}\, 3.4\!\cdot\!10^{16}$ & $4.5\!\cdot\! 10^{9} \text{--}\, 2.5\!\cdot\! 10^{10}$ & $7.8\!\cdot\! 10^{15} \text{--}\, 3.4\!\cdot\!10^{16}$ & $4.5\!\cdot\! 10^{9} \text{--}\, 1.6\!\cdot\! 10^{10}$ & \cmark\\
II7: & $4.4\!\cdot\! 10^{15} \text{--}\, 3.4\!\cdot\!10^{16}$ & $4.5\!\cdot\! 10^{9} \text{--}\, 7.2\!\cdot\! 10^{11}$ & $8.0\!\cdot\! 10^{15} \text{--}\, 3.4\!\cdot\!10^{16}$ & $4.5\!\cdot\! 10^{9} \text{--}\, 1.7\!\cdot\! 10^{11}$ & \xmark\\
II8: & $3.7\!\cdot\! 10^{15} \text{--}\, 4.7\!\cdot\!10^{15}$ & $1.1\!\cdot\! 10^{7} \text{--}\, 3.4\!\cdot\! 10^{9}$ & \multicolumn{2}{c|}{will be excluded} & \cmark \\\hline
III1: & $2.9\!\cdot\! 10^{14} \text{--}\, 3.4\!\cdot\!10^{16}$ & $4.5\!\cdot\! 10^{9} \text{--}\, 7.2\!\cdot\! 10^{11}$ & $3.0\!\cdot\! 10^{14} \text{--}\, 3.4\!\cdot\!10^{16}$ & $4.5\!\cdot\! 10^{9} \text{--}\, 1.7\!\cdot\! 10^{11}$ & \cmark\\
III2: & $M_X \gtrsim 2.3 \!\cdot\! 10^{15}$ & $M_1 \lesssim 1.8\!\cdot\! 10^{13}$ & $M_X \gtrsim 2.5 \!\cdot\! 10^{15}$ & $M_1 \lesssim 8.6\!\cdot\! 10^{12}$ & \cmark\\
III3: & $M_X \gtrsim 2.8 \!\cdot\! 10^{14}$ & $M_1 \lesssim 2.2\!\cdot\! 10^{12}$ & $M_X \gtrsim 2.8 \!\cdot\! 10^{14}$ & $M_1 \lesssim 9.5\!\cdot\! 10^{11}$ & \cmark\\
\multirow{2}{*}{III4:} & \multirow{2}{*}{$3.7\!\cdot\! 10^{15} \text{--}\, 6.1\!\cdot\!10^{15}$} & $1.6\!\cdot\! 10^{10} \text{--}\, 2.7\!\cdot\! 10^{10}$ & \multicolumn{2}{c|}{\multirow{2}{*}{will be excluded}} & \multirow{2}{*}{\cmark}\\
& & $\text{or}\; 8.7\!\cdot\! 10^{12} \text{--}\, 4.4\!\cdot\! 10^{13}\!\!\!$ & & &\\
III6: & $7.5\!\cdot\! 10^{14} \text{--}\, 4.8\!\cdot\! 10^{15}$ & $M_1 \lesssim 2.3\!\cdot\! 10^{10}$ & \multicolumn{2}{c|}{will be excluded} & \cmark\\
III7: & $4.5\!\cdot\! 10^{14} \text{--}\, 4.7\!\cdot\! 10^{15}$ & $M_1 \lesssim 6.4\!\cdot\! 10^{9}$ & \multicolumn{2}{c|}{will be excluded} & \cmark\\
III8: & $3.8\!\cdot\! 10^{15} \text{--}\, 3.6\!\cdot\! 10^{17}$ & $M_1 \lesssim 3.3\!\cdot\! 10^{9}$ & \multicolumn{2}{c|}{will be excluded} & \cmark\\
III10: & $1.0\!\cdot\! 10^{15} \text{--}\, 3.0\!\cdot\! 10^{17}$ & $M_1 \lesssim 3.0\!\cdot\! 10^{8}$ & $1.0\!\cdot\! 10^{15} \text{--}\, 3.0\!\cdot\! 10^{17}$ & $M_1 \lesssim 1.3\!\cdot\! 10^{8}$ & \cmark\\\hline
IV1: & $3.7 \!\cdot\! 10^{15} \text{--}\, 4.4 \!\cdot\! 10^{15}$ & $M_1 \lesssim 2.3\!\cdot\! 10^{10}$ & \multicolumn{2}{c|}{will be excluded} & \cmark\\
IV2: & $M_X \gtrsim 3.3 \!\cdot\! 10^{15}$ & $M_1 \lesssim 9.3\!\cdot\! 10^{12}$ & $M_X \gtrsim 6.0 \!\cdot\! 10^{15}$ & $M_1 \lesssim 3.8\!\cdot\! 10^{12}$ & \cmark\\
IV3: & $M_X \gtrsim 3.3 \!\cdot\! 10^{15}$ & $M_1 \lesssim 9.3\!\cdot\! 10^{12}$ & $M_X \gtrsim 6.0 \!\cdot\! 10^{15}$ & $M_1 \lesssim 3.8\!\cdot\! 10^{12}$ & \cmark\\
\hline\hline
\end{tabular}
}
\caption{Bounds on the GUT scale, $M_X$, and predicted ranges of the lowest intermediate scale, $M_1$, from Super-K and the future Hyper-K experiment if a null result of the proton decay is observed after a 10-year exposure. For each breaking chain, the gauge unification connects the intermediate scales with the GUT scale and gauge couplings. Super-K has set a lower bound on the proton lifetime $\tau_{\pi^0 e} > 1.6 \times 10^{34}$ years and Hyper-K is expected to set a bound $\tau_{\pi^0 e} > 1.4 \times 10^{35}$ years in the future. This information can be transformed to bounds on $M_X$ and further transformed to information of the prediction of any intermediate scales, e.g., $M_1$, using RG running and the constraint of gauge unification. In this Table, only breaking chains that have not been excluded are shown. The last column indicates if the chain generates observable GWs from cosmic strings. In order to generate such observable GWs, cosmic strings, but no other unnecessary topological defects should be generated from the last step of intermediate symmetry breaking.} \label{tab:pd_scales}
\end{table}

The RGE analysis described in \secref{sec:RGR}, provides not only the information of the scale of GUT symmetry breaking but also the information of scales of all intermediate symmetry breaking. Therefore, from our determination of $M_{1}$, which corresponds to the breaking of the lowest intermediate symmetry $G_1$, we can find the scale of cosmic string formation and calculate the associated stochastic gravitational wave background (SGWB). In this Section, we present the general formulation for calculating the 
gravitational wave spectrum generated by cosmic strings with string tension, $\mu$, and discuss the
various experiments which will have sensitivity to such signals. In \secref{sec:stringtension}, we connect the 
string tension with the lowest intermediate scale, $M_1$, which has been calculated for each type (c) breaking chain
in the earlier sections.

The network of cosmic strings formed during the breaking of the GUT to SM gauge groups acts as a source of GWs produced when the cosmic strings intersect to form loops. 
Cusps on these strings emit strong beams of 
high-frequency GWs or {\it bursts}. Furthermore, loops oscillate, shrink and emit energy gravitationally. This gravitational radiation constitutes a SGWB if they are unresolved over time \cite{Damour:2001bk,Damour:2004kw}.
 We assume a standard cosmology and that inflation occurs before string formation, and hence an undiluted GW spectrum may be observed\footnote{We note that non-standard cosmologies may affect the GW spectrum associated with the decay of cosmic strings \cite{Gouttenoire:2019kij}.}. 

To compute the SGWB, we follow the approach of \cite{Cui:2018rwi} where we have assumed Nambu-Goto strings that predominantly decay via gravitational radiation. For cosmic strings generated from gauge symmetry breaking, typical in GUTs, the energy released from the string decay may be transferred to gravitational radiation and into excitations of their elementary constituents. However, it has been shown that in the absence of long-range interactions, massive excitations of the vacuum (which is the case for GUTs) are suppressed for long-wavelength modes of the strings \cite{Auclair:2019wcv}. Furthermore, simulations of individual strings from the Abelian Higgs model show that particle production  is mainly important for small loops, and therefore the gravitational wave production is dominant for large loops \cite{Matsunami:2019fss}. 
As such, we assume there is no qualitative change for strings from gauge symmetry breaking. However, large-scale field theory simulations of the whole network of strings show discrepancies with this statement \cite{Hindmarsh:2017qff, Hindmarsh:2021mnl}. They show loops formed by infinite strings from random-field initial conditions can decay quickly. Due to this discrepancy, there may be significant uncertainties associated with the constraints on the cosmic string scale \cite{Hindmarsh:2017qff}. We anticipate this issue will be clarified in the coming years before the next-generation neutrino and GW experiments start data taking.

For the Nambu-Goto strings, the large loops provide the dominant contribution to the GW signal, and therefore we focus on them. The initial large loops have typical length $l_i = \alpha t_i$ with $\alpha \simeq 0.1$ which has been obtained numerically \cite{Blanco-Pillado:2013qja,Blanco-Pillado:2017oxo} and $t_i$ the initial time of string formation. 
The length of loops decreases as they release energy to the cosmological background, 
\begin{eqnarray} \label{eq:length}
l(t) = l_i - \Gamma G \mu (t-t_i)
 \,.
\end{eqnarray}
A loop of length $l$ oscillating in its $k$th harmonic excitation (for $k=1,2,\cdots$) will emit GWs of a frequency $2k/l$ in the early Universe. This radiation is  then redshifted down to its present-day value due to the Hubble expansion.
After the strings form, loops are found to emit energy in the form of gravitational radiation at a constant rate:
\begin{eqnarray}
\frac{dE}{dt} = - \Gamma G \mu^2 \,,
\end{eqnarray}
where $\Gamma \approx 50$ \cite{Burden:1985md,Vilenkin:2000jqa,Blanco-Pillado:2017oxo}. 
Assuming that the fraction of the energy transfer is in the form of large loops is $\mathcal{F}_\alpha \simeq 0.1$, the relic GW density parameter is given by
\begin{eqnarray}
\Omega_{\rm GW} (f) = \frac{1}{\rho_c} \frac{d\rho_{\rm GW}}{d \log f}.
\end{eqnarray}
This can be written as a sum of mode $k$
\begin{eqnarray}
\Omega_{\rm GW}(f) = \sum_k \Omega_{\rm GW}^{(k)}(f),
\end{eqnarray}
with 
\begin{eqnarray}
\Omega_{\rm GW}^{(k)} (f) &= &\frac{1}{\rho_c} \frac{2k}{f} \frac{\mathcal{F}_\alpha \Gamma^{(k)} G \mu^2}{\alpha (\alpha + \Gamma G \mu)} 
\int^{t_0}_{t_F} dt \frac{C_{\rm eff} (t_i^{(k)})}{t_i^{(k)4}} \frac{a^2(t) a^3(t_i^{(k)})}{ a^5(t_0)} \theta(t_i^{(k)} - t_F),
\label{OmegaGW}
\end{eqnarray}
where $\rho_c$ is the critical energy density of the Universe given by
\begin{eqnarray}
\Gamma^{(k)} &=& \frac{1}{3.6}\Gamma k^{-4/3}\\
t_i^{(k)} &=& \frac{1}{\alpha + \Gamma G \mu} \left( \frac{2k}{f} \frac{a(t)}{a(t_0)} + \Gamma G \mu t \right).
\end{eqnarray}
 $C_{\rm eff} = 5.7, 0.5$ has been numerically calculated \cite{BlancoPillado:2011dq,Blanco-Pillado:2017oxo,Blanco-Pillado:2013qja} for radiation and matter domination, respectively, and $t_F$ is the time of string network formation. 
In our numerical calculation of the SGWB, we have fixed the numerical values of $\alpha$ and $\mathcal{F}_\alpha$ at values suggested by simulation in the literature.

Applying these standard assumptions, we recover the general behaviour of SGWB spectrum from a cosmic string network: The GW spectrum peaks at low frequency and forms a flat plateau at high frequency, which refer to GWs emission during the matter domination and radiation domination eras, respectively, which has been discussed in detailed, e.g. in \cite{Figueroa:2020lvo}. In particular, we confirm an important feature of this kind of GWs that the amplitude of the GW spectrum at the flat plateau is proportional to $(G\mu)^{1/2}$ \cite{Auclair:2019wcv}.

A large range of $G\mu$ values can be explored using current and future GW detectors. LIGO O3 \cite{LIGOScientific:2019vic} has excluded cosmic strings formation at $G\mu \sim 10^{-8}$ in the high frequency range $10$-$100$~Hz. While in the nanoHertz regime, the null result of EPTA \cite{Lentati:2015qwp} and NANOGrav 11-year data \cite{Arzoumanian:2018saf} constrains the upper bound of $G\mu \lesssim 6\times 10^{-11}$. 
The strongest constraint is provided by PPTA collaboration, $G\mu \lesssim 1.5 \times 10^{-11}$ \cite{Blanco-Pillado:2017rnf}.
In \cite{Arzoumanian:2018saf}, the NANOGrav Collaboration presented its search results for an isotropic SGWB based on its 12.5-year data set.  The source of this signal could be astrophysical, however, it could possibly
be an indication of cosmic strings.

 Planned pulsar timing arrays SKA \cite{Janssen:2014dka}, space-based laser interferometers LISA \cite{Audley:2017drz}, Taiji \cite{Guo:2018npi}, TianQin \cite{Luo:2015ght}, BBO \cite{Corbin:2005ny}, DECIGO \cite{Seto:2001qf}, ground-based interferometers Einstein Telescope \cite{Sathyaprakash:2012jk} (ET), Cosmic Explorer \cite{Evans:2016mbw} (CE), and atomic interferometers MAGIS \cite{Graham:2017pmn}, AEDGE \cite{Bertoldi:2019tck}, AION \cite{Badurina:2019hst} will probe $G \mu$ values in a wide range $\sim 10^{-19} - 10^{-11}$. Finally, it was recently highlighted \cite{Garcia-Bellido:2021zgu} that
large surveys of stars such as Gaia \cite{Brown:2018dum} and the proposed
upgrade, THEIA \cite{Boehm:2017wie}, can be powerful probes of gravitational waves (GW) in the same frequency regime as SKA.

\section{The interplay between proton decay and gravitational waves in $SO(10)$ GUTs}\label{sec:stringtension}
This Section connects the constraints placed on $SO(10)$ GUTs by current and future proton decay limits to infer
the gravitational wave signature associated with each breaking chain. In \secref{sec:stringandmu}, we explain how the lowest intermediate scale is connected to the string tension, and in \secref{sec:testability} we assess the testability of these various chains in light of current and upcoming gravitational wave detectors sensitivities.

\subsection{String tension bounded by proton decay}\label{sec:stringandmu}
Having discussed the sensitivities of various GW detectors to $G\mu$, we presently provide a connection between this
phenomenological string parameter, $\mu$, and the lowest intermediate scale, $M_1$, the string formation scale.
In the paradigm of gauge symmetry breaking, the tension of the cosmic strings is correlated with the Higgs VEV and the symmetry breaking scale. 
Moreover, in the simplest case with just a complex Higgs $\phi$ and the breaking of a single $U(1)$ gauge symmetry, the string tension is given by $\mu = 2\pi v^2 n \epsilon_n$, where the integer $n$ is the winding number of the vortex solution, $v = |\langle \phi \rangle|$ is the absolute value of the Higgs VEV, and $\epsilon_n$ is an ${\cal O}(1)$ function weakly depending on the mass-squared ratio of the Higgs to the gauge boson $\beta = m_\phi^2/m_{Z'}^2$ \cite{Hindmarsh:1994re}. In all cases, $n=1$ provides a topologically stable string. Numerical results have  shown that $\epsilon_1$ is slowly increasing with $\beta$, in particular, limited in $0.5 < \epsilon_1 <3$ for a large region of the mass-squared ratio $0.01 <\beta < 100$,  $0.2 < \epsilon_1 <8$ for $10^{-6} < \beta < 10^{6}$, and $\epsilon_1 = 1$ when the Higgs mass equals the gauge boson mass \cite{Hill:1987qx}. Without knowing more details, one can not determine this ${\cal O}(1)$ factor quantitatively. Below, we assume $n=1$ strings dominate the string network and ignore the ${\cal O}(1)$ factor. Given the $U(1)'$ gauge boson mass $M_{Z'}^2 = 4\pi \alpha v^2$,
the string tension is correlated to the gauge boson mass scale via $G\mu \simeq (2 \alpha)^{-1} G M_{Z'}^2$.

\begin{figure}[t!]
\centering
\includegraphics[width=.75\textwidth]{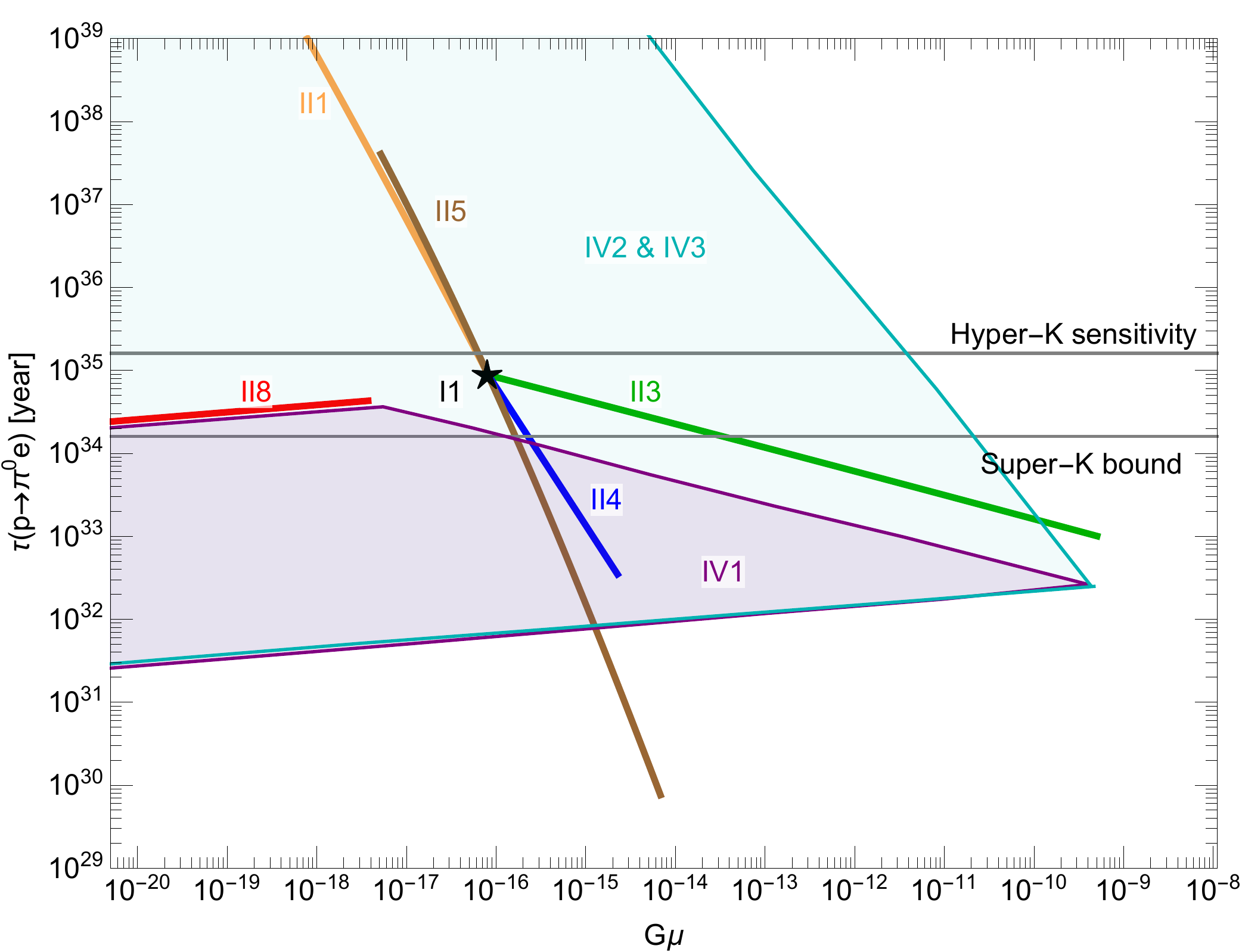}
\includegraphics[width=.75\textwidth]{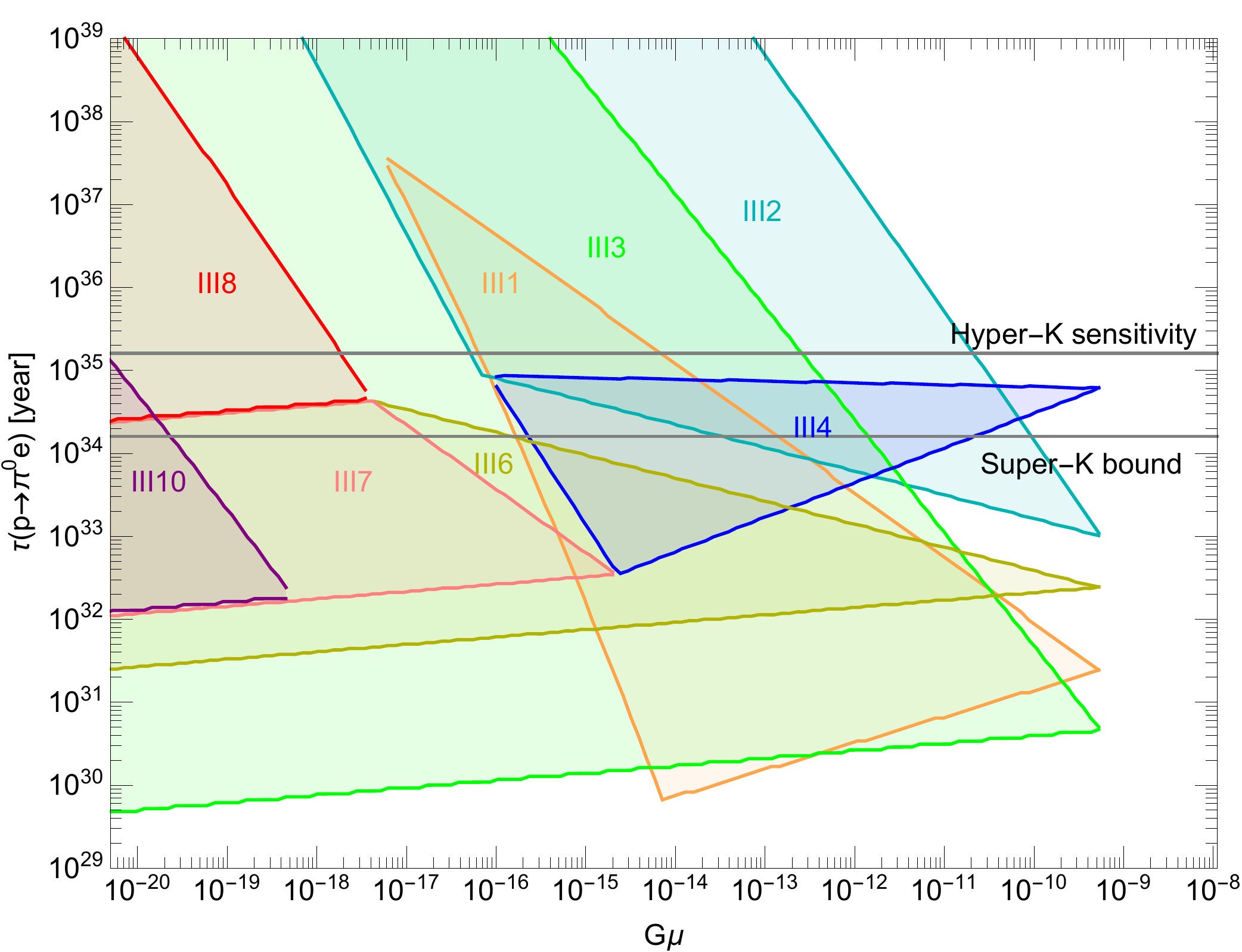}
\caption{Correlation between $G\mu$ and the proton decay lifetime for chains which can generate observable cosmic strings. Chains with one, two and four intermediate scales are shown in the upper panel and those with three intermediate scales are shown in the lower panel. 
The colour for each chain is specified in both the upper and low panel. Labels for each chains are not shown in the upper panel as regions for these chains are highly overlapped.}
\label{fig:Gmu}
\end{figure}

\begin{table}[t!]
\resizebox{\textwidth}{!}{%
\centering
\begin{tabular}{|c|c|c|}
\hline\hline
Chain & $G\mu$ compatible with Super-K & $G\mu$ after Hyper-K (no proton decay)\\\hline
I1 & $G\mu \simeq 2.0\times 10^{-17}$ & will be excluded \\\hline
II1: & $G\mu \lesssim 2.0 \times 10^{-17}$ & $G\mu \lesssim 1.5 \times 10^{-17}$ \\
II3: & $G\mu \simeq 2.0 \times 10^{-17} \text{--}\, 8.4 \times 10^{-15}$ & will be excluded \\
II4: & $G\mu \simeq 2.0 \times 10^{-17} \text{--}\, 5.5 \times 10^{-17}$ & will be excluded \\
II5: & $G\mu \simeq 1.2 \times 10^{-18} \text{--}\,4.0 \times 10^{-17}$ & $G\mu \simeq 5.1 \times 10^{-18} \text{--}\, 6.3 \times 10^{-17}$\\
II8: & $G\mu\lesssim 1.0 \times 10^{-18}$ & will be excluded \\\hline
III1: & $G\mu \simeq 1.3 \times 10^{-18} \text{--}\, 3.3 \times 10^{-14}$ & $G\mu \simeq 1.3 \times 10^{-18} \text{--}\, 1.6 \times 10^{-15}$ \\
III2: & $G\mu \lesssim2.3 \times 10^{-11}$ & $G\mu \lesssim5.0 \times 10^{-12}$ \\
III3: & $G\mu \lesssim3.4 \times 10^{-13}$ & $G\mu \lesssim6.2 \times 10^{-14}$ \\
III4: & $G\mu \simeq 2.1 \times 10^{-17} \text{--}\, 1.3 \times 10^{-10}$ & will be excluded \\
III6: & $G\mu \lesssim3.6 \times 10^{-17}$ & will be excluded \\
III7: & $G\mu \lesssim3.6 \times 10^{-18}$ & will be excluded \\
III8: & $G\mu \lesssim1.0 \times 10^{-18}$ & will be excluded \\
III10: & $G\mu \lesssim5.6 \times 10^{-21}$ & $G\mu \lesssim 1.1 \times 10^{-21}$ \\\hline
IV1: & $G\mu \lesssim 3.1 \times 10^{-17}$ & will be excluded \\
IV2: & $G\mu \lesssim 5.1 \times 10^{-12}$ & $G\mu \lesssim 9.4 \times 10^{-13}$ \\
IV3: & $G\mu \lesssim 5.1 \times 10^{-12}$ & $G\mu \lesssim 9.4 \times 10^{-13}$ \\
\hline\hline
\end{tabular}
\caption{\label{tab:Gmutab}Breaking chains which generates observable GWs from cosmic strings and the predicted $G\mu$ which are compatible with the Super-K bound and the future Hyper-K experiment limit on the proton decay lifetime (assuming no signal). Only breaking chains which have not been excluded are shown. For each breaking chain, the gauge unification connects intermediate scales with the GUT scale and gauge couplings. Super-K has set a lower bound on the proton life time $\tau_{\pi^0 e} > 1.6 \times 10^{34}$ years and Hyper-K is expected to set a bound $\tau_{\pi^0 e} > 1.4 \times 10^{35}$ years, assuming no signal of proton decay is observed. This information can be transformed bounds on $G\mu$, with RG running and gauge unification taken into account.
 \label{tab:Gmu}}
}
\end{table}

Let us first consider strings generated from the breaking: $G_{3211} \to G_{\rm SM}$, i.e., $U(1)_R \times U(1)_X \to U(1)_Y$. The lowest intermediate scale $M_1$, defined to be the gauge boson mass associated with this symmetry breaking,
is given by $M_1^2 \simeq 4 \pi (\alpha_{1R}(M_1)+\alpha_{1X}(M_1)) v^2$, where $\alpha_{1R}(M_1)$ and $\alpha_{1X}(M_1)$ are gauge coefficients of $U(1)_R$ and $U(1)_X$ fixed at the scale $M_1$, respectively. Therefore, we can approximate 
\begin{eqnarray}
G\mu \simeq \frac{1}{2(\alpha_{1R}(M_1)+\alpha_{1X}(M_1))} \frac{M_1^2}{M_{\rm pl}^2} \,.
\end{eqnarray}
The direct breaking from a non-Abelian symmetries such as $G_{3221}$ and $G_{421}$ to $G_{\rm SM}$ can also generate observable cosmic strings without the production of any unnecessary defects. The tension of strings generated via $G_{3221} \to G_{\rm SM}$ and $G_{421} \to G_{\rm SM}$ is approximatively given by
\begin{eqnarray}
G\mu \simeq \left\{
\begin{array}{c}
\displaystyle\frac{1}{2(\alpha_{2R}(M_1)+\alpha_{1X}(M_1))} \frac{M_1^2}{M_{\rm pl}^2}\,, \\
\\
\displaystyle\frac{1}{2(\alpha_{4c}(M_1)+\alpha_{1R}(M_1))} \frac{M_1^2}{M_{\rm pl}^2}\,,
\end{array}
\right.
\end{eqnarray}
where $\alpha_{2R}(M_1)$ and $\alpha_{4c}(M_1)$ is the gauge coefficient of $SU(2)_R$ and $SU(4)_c$ at $M_1$, respectively. The values of these gauge coefficients have been derived using the RGE analysis for each breaking chain. Moreover, we will ignore the small order one factor associated with the one-loop matching condition
and estimate this as an order one uncertainty. At this present stage, this level of uncertainty is acceptable given the experimental and string simulation uncertainties.
The direct breaking from $G_{3221}^C$, $G_{422}$ or $G_{422}^C$ to $G_{\rm SM}$ generates both strings and unnecessary topological defects (either domain walls or monopoles), and one has to include inflation to dilute these unwanted defects. Consequently, the string network is completed diluted, and no GW signal is generated. 

\subsection{Testability of $SO(10)$ via gravitational waves}\label{sec:testability}
After considering all possible non-supersymmetric $SO(10)$ breaking chains which provide gauge unification, can generate a GW signal and are not excluded by the current proton decay constraints from Super-K, the only remaining chains are listed in Table~\ref{tab:Gmu}. In that Table, the values of $G\mu$ allowed by the potential null result from the future Hyper-K experiment are also listed. Notably, several of these chains will be excluded if a null result emerges from the future Hyper-K measurement. A more detailed correlation between $G\mu$ and the proton lifetime is shown in Fig.~\ref{fig:Gmu}. For example, if Hyper-K does not observe proton decay, then I1, II3, II4 and II8 would be excluded, and the viable chains with two intermediate scales or less would be II1 and II5. Similarly, chains IV1, III4, III6, III7 and III8 would be excluded by a null proton decay result from Hyper-K.

\begin{figure}[t]
\centering
\includegraphics[width=0.74\textwidth]{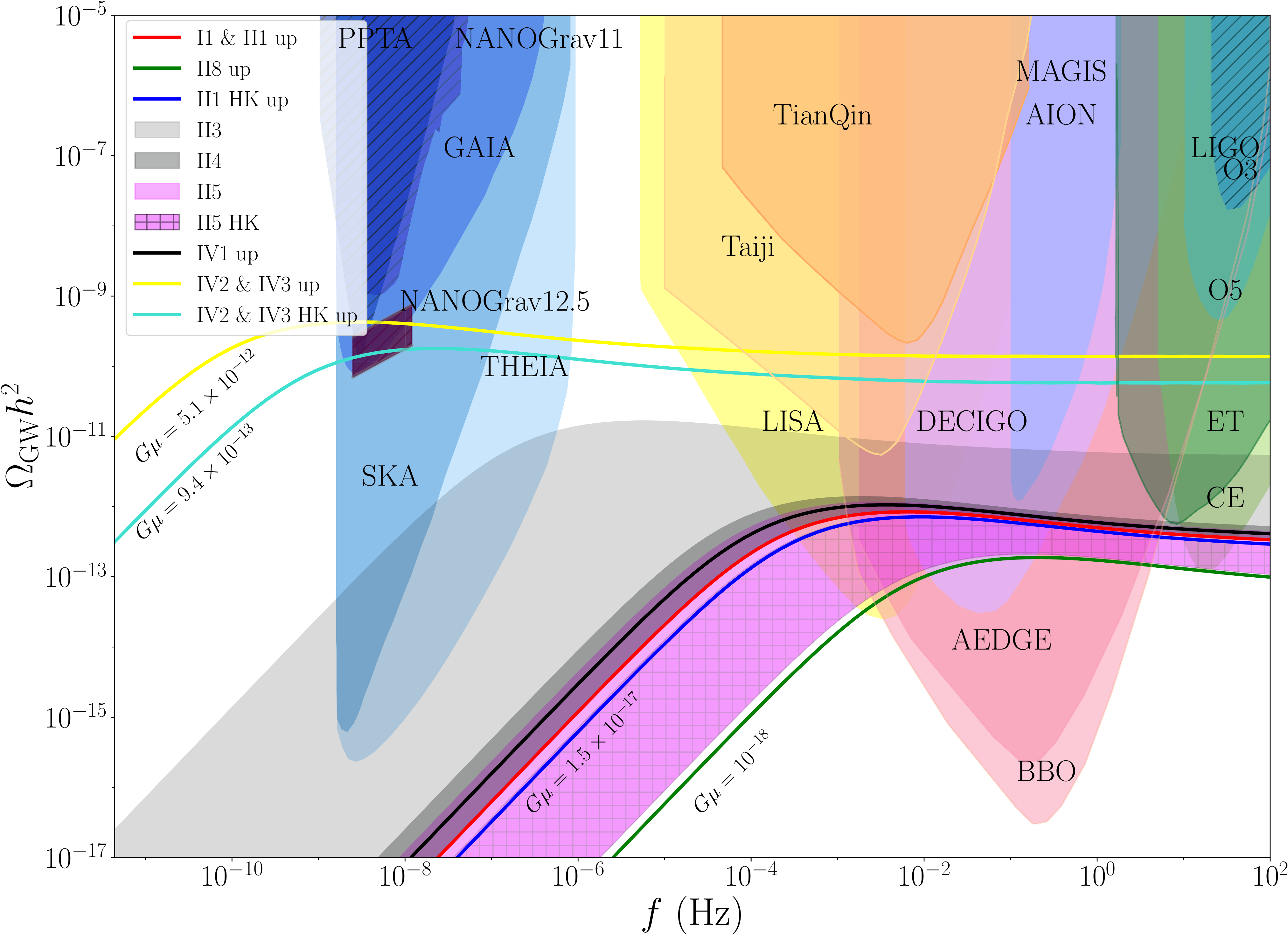}
\includegraphics[width=0.74\textwidth]{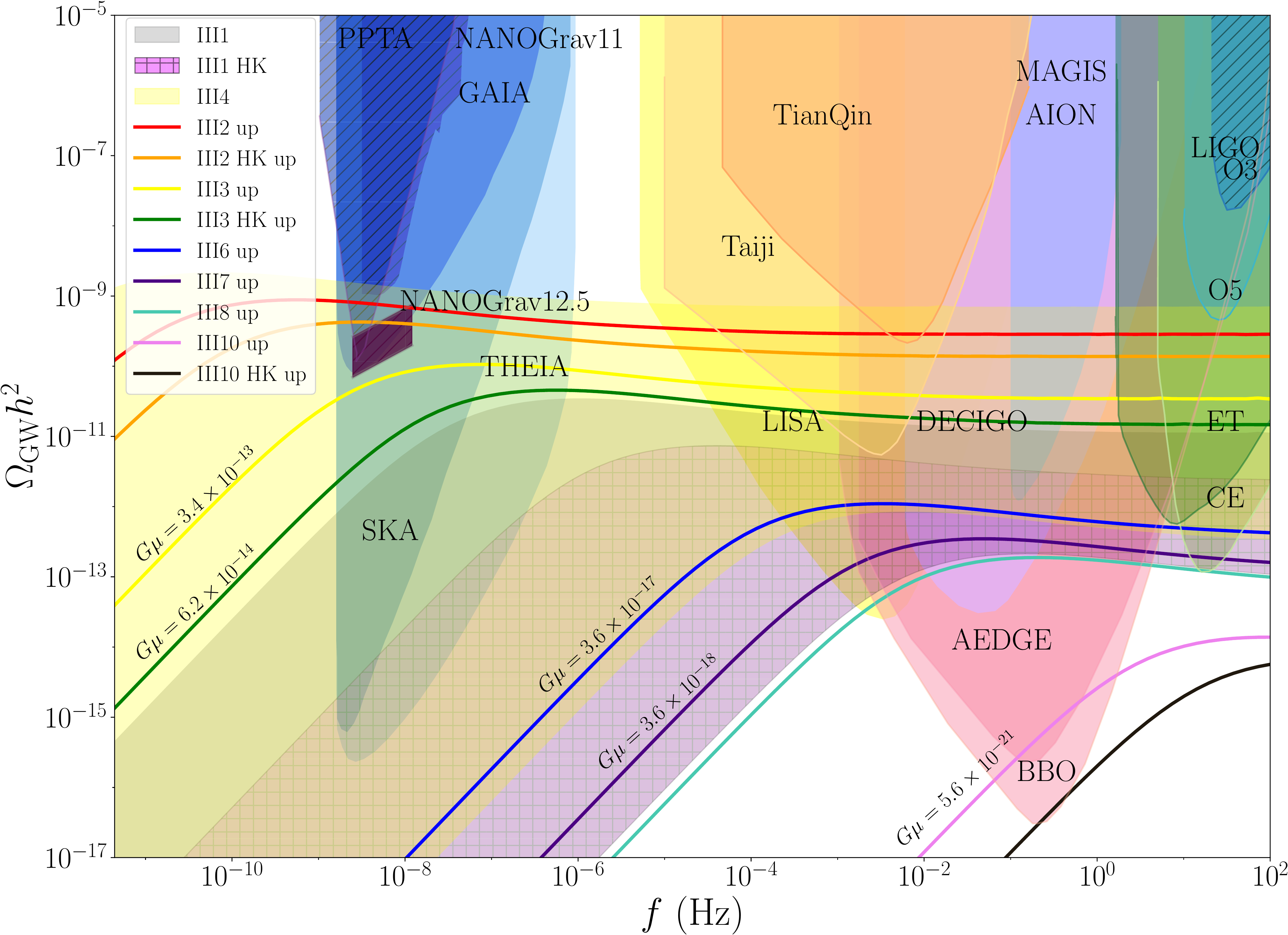}
\caption{SGWB predicted from cosmic string networks shown in Table.~\ref{tab:Gmu}. The solid coloured lines
indicate signal or upper bounds (if it is an upper bound, this is indicated as ``up'' in the legend) for various type (c) breaking chains. The solid (hatched) coloured regions indicate the range of values of $G\mu$ predicted by the non-observation of proton decay by Super-K (Hyper-K). 
The SGWB for II3, II4 and II5 consistent with the current Super-K bound are shown in light grey, dark grey, and pink. The yellow hatched region shows the SGWB for II5 consistent with the non-observation of proton decay at Hyper-K. Current (future) experimental limits are shown by hatched (solid) regions. The dark purple rhomboidal region shows the signal region observed by NANOGrav 12.5.} 
\label{fig:GW1}
\end{figure}

In the following discussion, we assume that the only source of GWs  comes from GUT symmetry breaking.
The resulting sensitivities of the various GW detectors and the GW signal from type-I, II, IV and III chains are shown in
upper and lower plots of \figref{fig:GW1}, respectively. The solid coloured lines indicate the GW signal or the upper bound on the signal (``up'' in the legend of \figref{fig:GW1} refers to an upper bound, and the absence of ``up'' indicates a single-valued signal). The block (hatched) coloured regions show the range of $G\mu$ consistent with the Super-K (prospective Hyper-K) bound for each chain. In the upper plot, we observe that the current bound on proton decay constrains chains II3, II4 and II5 to have GW signals in the region of parameter space tested by LISA, Taiji, MAGIS AION, DECIGO, AEDGE, BBO and CE. However, if Hyper-K does not observe proton decay, then I1, II3, II4 and II8 would have no GW signal associated with those chains. Naturally, the null observation of proton decay and the observation of GWs would exclude these chains. However, the prospective null-observation by Hyper-K would constrain II5 to have a GW signal detectable by LISA, AEDGE, BBO and CE. Interestingly, the only chain with two or less intermediate symmetries, which would be viable with the possibility of dual non-observation of proton decay and GW is II1 as the associated SGWB could have a very small amplitude undetectable even by BBO. Nonetheless, the null result of Hyper-K and the observation of a GW signal with $G\mu\gtrsim 10^{-17}$ would exclude II1.
Finally, IV1 has no associated GW signal for this chain given a null-observation by Hyper-K; however, IV2 and IV3  remain viable with only an upper bound on their associated GW signal as indicated by cyan on the upper plot of \figref{fig:GW1}. If a GW signal was observed, alongside non-observation of proton decay by Hyper-K, with 
$G\mu\gtrsim 10^{-12}$, then IV2 and IV3 would be excluded. 
In the lower plot, we observe that the current bounds on III1 and III4 can be tested by many experiments which probe the lower and high-frequency
regimes. The constraint on the remaining chains, from the current bound on proton decay, simply places an upper bound on their GW
signal. Similarly to the discussion above, in the event of non-observation of proton decay by Hyper-K and an observation of a GW signal above any of the upper bounds, it would exclude the chains above.
In the scenario that Hyper-K does not observe proton decay, then there would be no GW signal associated with 
III4, III6, III7 and III8. An upper bound is placed on III2, III3 and III10, while III1 is constrained to have a GW signal in the region probed by 
LISA, DECIGO, AEDGE, BBO, ET and CE. The only chains which survive dual non-observation are, therefore, III2, III3 and III10. 

\subsection{Implications for Leptogenesis}

The unavoidable presence of the right-handed neutrinos in the $SO(10)$ theories explains the smallness of active neutrino masses via the type-I
seesaw mechanism \cite{Minkowski:1977sc,GellMann:1980vs,Yanagida:1979as,Mohapatra:1979ia}.
Furthermore, the out-of-equilibrium and CP-violating decays of these right-handed neutrinos can be generate
the matter-antimatter asymmetry via thermal type-I leptogenesis \cite{Fukugita:1986hr}.
Such a framework, which simultaneously
provides gauge unification, a solution to the flavour problem, an explanation of neutrino masses
and the matter-antimatter asymmetry is highly attractive and has been studied extensively \cite{Nezri:2000pb,DiBari:2005st,DiBari:2010ux,Buccella:2012kc,DiBari:2015oca,DiBari:2020plh}.

As pointed out in \cite{Dror:2019syi}, the $U(1)$ breaking connected with leptogenesis can be associated with a SGWB. In our models, the $U(1)$ symmetry breaking that generates the cosmic strings, and hence the possible GW signature
is not a free parameter but corresponds to the breaking of a gauged ${B-L}$ symmetry that forbids right-handed neutrino masses. As discussed, the scale $M_1$ is not free but depends on the specific breaking chain and is constrained by gauge coupling unification and proton decay limits. This scale corresponds to the highest possible mass of the right-handed neutrinos, $N_3$, assuming perturbative Yukawa couplings, such that $M_{N_3}\lesssim M_{N_1}$.
The parameter relevant for leptogenesis is the mass of the lightest right-handed neutrino $M_{N_1}$ or, if leptogenesis is controlled by the decay of the second-lightest neutrino $N_2$, $M_{N_2}$. Therefore, $M_1$ constrains also the leptogenesis scale, the latter typically being much lower than $M_1$, so that $M_{N_{1,2}} \ll M_{N_{3}}  \lesssim M_1 $. 

Proton decay and GW signatures or lack of thereof, will therefore set an upper limit (or possibly a range) for $M_1$ and hence
the heaviest right-handed neutrino mass. Already tension is present for many breaking chains. Successful thermal type-I non-resonant leptogenesis requires at least one right-handed neutrino to have a mass
of order $10^{10}$ GeV, implying $M_1 \gg 10^{10}$ GeV. For instance, chains I4, II8, III7, III8, III10 already appear in tension with the most naive implementation of the leptogenesis scenario, and these bounds will be critically strengthened by the future HK results.

For example, in one commonly assumed
 $SO(10)$-inspired  scenario \cite{DiBari:2020plh}, the right-handed
neutrino mass spectrum follows the pattern of mass splittings in the up-type quark sector which provides a strongly hierarchical spectrum, $M_{N_1} : M_{N_2}: M_{N_3} \sim m_u^2 :m_c^2 : m_t^2$. In this case, the mass of the lightest right-handed neutrino is too small to generate a sufficient lepton asymmetry, 
and so the second heaviest right-handed neutrinos with a mass of order $10^{10}$ GeV
are used to create the observed matter-antimatter asymmetry \cite{DiBari:2005st,DiBari:2020plh}. 
Even so, $M_{N_2} \sim 10^9-10^{10}$~GeV implies $M_{N_3}$ must be at least $10^{13}$~GeV, which implies that only chains III2, III4 and possibly IV2 and IV3 are viable. Such chains would induce a GW signal within reach at current and next generation of experiments, for instance, the tantalising results of NANOGrav. A definitive non-observation of GW corresponding at such scale, together with the assumption of a high inflationary scale, would place high-scale  type-I thermal leptogenesis under significant tension as the origin of the baryon asymmetry of the universe, at least in its simplest implementation of $SO(10)$. 

With the prospect of HK results and future GW observations, similar but more compelling considerations could be drawn. If BBO did not observe a GW signal from cosmic strings then, provided the inflationary scale was high, 
the upper limit intermediate scale breaking would be $M_N \sim 10^9$ GeV, which would prevent any right-handed neutrinos from gaining masses higher than this scale and preclude this scenario.
This conclusion would also impact concrete $SO(10)$ models
where the right-handed neutrinos are hierarchical with the heaviest right-handed neutrino mass larger than $10^9$ GeV \cite{King:2003rf}.
In general, an upper limit of $10^9$ GeV on the ${B-L}$ symmetry breaking scale would imply that any model based on
non-resonant type-I thermal leptogenesis would be difficult to achieve without fine-tuning of the Yukawa matrix \cite{Moffat:2018wke}. 
Indeed, the requirement of a GUT scale right-handed neutrino mass seems at odds with the conditions for unification which generically requires that $M_1<M_{X}$, at least in the non-supersymmetric case considered here.

The consistency of the leptogenesis parameter space, in the frameworks of GUTs, the link with light neutrino masses and the dependence on the hierarchy of right-handed neutrino masses as well as on the assumption controlling the running of the gauge couplings, in light of the next generation proton decay and GW limits, is of particular interest. Such examples, together with type II leptogenesis, will be considered in future work.

\section{Summary and Conclusions}\label{sec:sumanddis}
We have considered all possible breaking chains of $SO(10)$ and determined which ones
can be dually tested by proton decay and gravitational waves experiments. We found that type (c)
chains, namely chains with the Pati-Salam group as an intermediate symmetry, can provide unification without supersymmetry and produce a 
GW signal. From our renormalisation group equation analysis
we found that only 17 of the 31 type (c) breaking chains have not already been excluded by current proton decay limits. Given the null observation of proton decay at Super-K and the prospective null observation at Hyper-K, we determined bounds on the GW signal of each of these 17 chains. We found that:
\begin{itemize}
\item if Hyper-K does not observe proton decay, then nine of the 17 type (c) breaking chains (I1, II3, II4, II8, III4, III6, III7, III8 and IV1) can be excluded and, as such, no GW signal would be associated to these chains. 
\item Of the remaining eight type (c) chains, the non-observation of proton decay by Hyper-K sets upper limits on the GW signal of six of these chains (II1, III2, III3, III10, IV2 and IV3). 
This implies that for a given chain, the non-observation of proton decay combined with the observation of a GW signal above the upper limit of that chain would exclude that particular chain. The upper limits of $G\mu$ for these chains are given in Table~\ref{tab:Gmu} and shown in \figref{fig:GW1}.

\item In the event of non-observation of proton decay by Hyper-K, this constrains
II5 and III1 to have a range of $G\mu$ values and therefore a definite GW signal. Both chains can be tested by a range of experiments sensitive to the higher frequency regime (these include LISA, DECIGO, AEDGE, CE, ET, MAGIS AION). 
Therefore, the non-observation of proton decay by Hyper-K combined with the observation of a GW signal in relevant ranges would positively indicate those chains.
\item In the exciting event that Hyper-K observes proton decay, then the measured proton decay lifetime will determine the intermediate scale, $M_1$, which will, in turn, have an associated value of $G\mu$, and therefore a corresponding GW signal, for each of the 17 type (c) breaking chains. Depending on the value of the proton lifetime, the associated GW signal will differ. Nonetheless, the procedure we outline in this paper can be used to assess each chain separately
and correlate the proton lifetime with the gravitational wave signal.
\end{itemize}

In our analysis, we have applied the result of simulations based on Nambu-Goto strings to strings from gauge symmetry breaking. It certainly induces several uncertainties to constraints of GW measurements on GUT intermediate scales. The first is the order-one factor from the string tension $\mu$ in terms of the lowest intermediate symmetry breaking scale $M_1$. The string tension may be slightly dependent upon the Higgs to gauge boson mass ratio, which is unknown until added. Another uncertainty is from the string network simulation. Until now, most simulations of SGWB from the string network are based on the Nambu-Goto strings, an approximation of the infinitely thin strings and no couplings to particles. It is supported by the simulation of individual strings in Abelian-Higgs theory, but not supported by the large-scale field theory simulation. These uncertainties can weaken the constraints on the lowest intermediate scale $M_1$ in GUTs. Given the timeline of future neutrino and GW experiments, we expect these uncertainties can be better controlled, leading to more concrete constraints before the next-generation experiments take data.

Finally, we would like to comment on a possible recent detection of SGWB by the NANOGrav collaboration. In \cite{Arzoumanian:2018saf}, the NANOGrav Collaboration presented its search results for an isotropic SGWB based on its 12.5-year data set. Interestingly, this study might yield
an indication for the presence of a SGWB
across the 45 pulsars included in their analysis.
GW from cosmic strings is a possible explanation of such signal \cite{Blasi:2020mfx,Ellis:2020ena,Buchmuller:2020lbh,Lazarides:2021uxv} with a value of $G\mu \sim 10^{-12}-10^{-11}$. This corresponds to a scale $M_1 \sim 10^{13}\,\text{GeV}$. Applying this interpretation, combined with the possible non-observation of proton decay by Hyper-K, then III4, III2, IV2 and IV3 could generate a signal in the region detected by NANOGrav12.5.

\section*{Acknowledgement}
This work was partially supported by the European Union’s Horizon 2020 Research and Innovation Programme under Marie Sklodowska-Curie grant agreement HIDDeN European ITN
project (H2020-MSCA-ITN-2019//860881-HIDDeN),  the European Research Council under ERC Grant NuMass (FP7-IDEAS-ERC ERC-CG 617143). S. F. K. acknowledges the STFC Consolidated Grant ST/T000775/1.

\newpage
\appendix
\section{$\beta$ coefficients at intermediate scales of GUT \label{app:beta_coefficients}}

This appendix lists $\beta$ coefficients of two-loop RG running functions used in our paper. 

For the scale vary from the electroweak scale to the scale $G_1$, the one-loop and two-loop $\beta$ coefficients are well-known, 
\begin{eqnarray}
\{b_i\} = \begin{pmatrix} -7 \\ -\frac{19}{6} \\ \frac{41}{10} \end{pmatrix} \,,\quad
\{b_{ij} \} = \begin{pmatrix} 
-26 & \frac{9}{2} & \frac{11}{10} \\
12 & \frac{35}{6} & \frac{9}{10} \\
\frac{44}{5} & \frac{17}{10} & \frac{199}{50} \\
\end{pmatrix}\,.
\end{eqnarray} 
These coefficients are obtained by including only gauge symmetries, SM fermions and the SM Higgs. 

Below, we discuss the deviation of $\beta$-coefficients at intermediate scales between the electroweak scale and the GUT scale. 
\begin{itemize}
\item
At these scales, intermediate symmetries $G_I \supset G_{\rm SM}$ are recovered. The SM Higgs may not be a doublet anymore but be embedded in a larger multiplet. To generate correct fermion mass spectrum, the SM Higgs should be combination of ${\bf 10}$ and $\overline{\bf 126}$ of $SO(10)$. 

\item
For breaking the intermediate symmetry $G_I$ to a smaller group, additional heavy Higgses are required to achieve the breaking. These Higgses contribute to the RG running from the scale of $G_I$ to any higher scales. Typical $SO(10)$ Higgs multiplets used to achieve these breakings are listed in Table~\ref{tab:chains_2}. 
For example, the $\overline{\bf 126}$ Higgs includes a $(\1, \1, 3\, -1)$ of $G_{3221}$ which further include a trivial singlet of $G_{\rm SM}$. Once the trivial singlet gains the VEV, $G_{3221}$ is broken to $G_{\rm SM}$. Contribution of this $(\1, \1, \3, -1)$ should be considered in the RG running from $G_{3221}$ to any larger symmetries. In an alternative breaking $G_{3221}^C \to G_{\rm SM}$, $(\1, \1, 3\, -1)$ can be also introduced to achieved the breaking, but an additional $(\1, \3, \1, +1)$ is required due to the left-right parity symmetry. Thus, both fields should be included in the running from $G_{3221}^C$. 
The parts of the Higgs GUT multiplets not required for symmetry breaking at lower scales are assumed to be heavy and decoupled at the higher scales, by some unspecified Higgs potential, which we assume not to introduce further Higgs multiplets which could affect the running.
\end{itemize}
The set of Higgs multiplets required for the considered symmetry breaking pattern may not be unique for the breaking of some intermediate symmetries. We consider the most economical case that minimal particle contents with all the above ingredients included. Once the gauge group and particle content are fixed, the $\beta$-coefficients are determined. $\beta$ coefficients in all breaking chains and all intermediate scales are listed in Table~\ref{tab:beta_coefficients}, with particle contents listed explicitly. All $\beta$-coefficients at intermediate scales for type (c) breaking chains are listed in Table~\ref{tab:beta_coefficients}. Some $\beta$ coefficients may be different from some chains appearing in \cite{Bertolini:2009qj, Chakrabortty:2019fov} since different Higgs fields are assigned. We have checked that our $\beta$ coefficients are the same as those in the one intermediate scale case in \cite{Meloni:2019jcf}. 


\begin{table}[H]
\caption{Coefficients of $\beta$ coefficients at intermediate scales with intermediate symmetries preserved for GUTs broken to the SM, $SO(10) \to \cdots \to G_2 \to G_1 \to G_{\rm SM}$. 
The matter field is arranged as ${\bf 16}$ of $SO(10)$ in all breaking chains, as indicated as ${\bf 16}_F$. The SM Higgs is considered to be embedded in a linear combination of ${\bf 10}$ and $\overline{\bf 126}$ of $SO(10)$, which are necessary to generate correct fermion mass spectrum. Extra Higgs have to be introduced to achieve the breaking. We include minimal Higgs contents for each intermediate symmetry breaking and list all matter and Higgs field in the table. \\}
\label{tab:beta_coefficients}
\centering
\begin{tabular}{|p{10mm}|p{57mm}|c|c| p{25mm}|}
\hline
\multirow{2}{*}{Symm.} & \multirow{2}{*}{Particle content} & \multicolumn{2}{c|}{$\beta$-coefficients} & \multirow{2}{*}{Applied to} \\\cline{3-4}
 & & $\{ b_i \}$ & $\{ b_{ij} \}$ & \\\hline
\multirow{4}{*}{$G_{3221}$} & 
{\footnotesize $\begin{array}{l}(\3, \2, \1, \pm\frac{1}{6})+ (\1, \2, \1, \pm\frac{1}{2})\subset {\bf 16}_F, \\
(\1, \2, \2, 0)_1\subset {\bf 10}_H,\\
(\1, \2, \2, 0)_2 + (\1, \1, \3, -1) \subset \overline{\bf 126}_H \end{array}$} &
{\footnotesize $\begin{pmatrix}-7 \\ -\frac{8}{3} \\ -2 \\ \frac{11}{2}\end{pmatrix}$} &
{\footnotesize $\begin{pmatrix}
 -26 & \frac{9}{2} & \frac{9}{2} & \frac{1}{2} \\
 12 & \frac{37}{3} & 6 & \frac{3}{2} \\
 12 & 6 & 31 & \frac{27}{2} \\
 4 & \frac{9}{2} & \frac{81}{2} & \frac{61}{2} \\
\end{pmatrix}$} &
$\begin{array}{l}G_1 \text{ in I1, } \\ \text{II1,3,4, III2,4} \end{array}$
 \\\cline{2-5}
& 
{\footnotesize $\begin{array}{l}(\3, \2, \1, \pm\frac{1}{6}) + (\1, \1, \2, \pm\frac{1}{2}) \subset {\bf 16}_F, \\
(\1, \2, \2, 0)_1\subset {\bf 10}_H,\\
(\1, \2, \2, 0)_2 + (\1, \1, \3, -1) \subset \overline{\bf 126}_H, \\ 
(\1, \1, \3, 0) \subset {\bf 45}_H 
\end{array}$} &
{\footnotesize $\begin{pmatrix} -7 \\ -\frac{8}{3} \\ -\frac{4}{3} \\ \frac{11}{2} \end{pmatrix}$} &
{\footnotesize $\begin{pmatrix}
 -26 & \frac{9}{2} & \frac{9}{2} & \frac{1}{2} \\
 12 & \frac{37}{3} & 6 & \frac{3}{2} \\
 12 & 6 & \frac{149}{3} & \frac{27}{2} \\
 4 & \frac{9}{2} & \frac{81}{2} & \frac{61}{2} \\
\end{pmatrix}$} &
$\begin{array}{l}G_2 \text{ in II8,} \\ \text{III6-8, IV1,2}\end{array}$
\\\hline
\multirow{7}{*}{$G_{3221}^C$} & 
{\footnotesize $\begin{array}{l}(\3, \2, \1, \pm\frac{1}{6})+ (\1, \2, \1, \pm\frac{1}{2}) \subset {\bf 16}_F, \\
(\1, \2, \2, 0)_1\subset {\bf 10}_H,\\
(\1, \2, \2, 0)_2+ (\1, \1, \3, \pm1) \subset \overline{\bf 126}_H
\end{array}$} &
\multirow{4}{*}{\footnotesize $\begin{pmatrix} -7 \\ -2 \\ -2 \\ 7 \end{pmatrix}$}&
\multirow{4}{*}{\footnotesize $\begin{pmatrix}
 -26 & \frac{9}{2} & \frac{9}{2} & \frac{1}{2} \\
 12 & 31 & 6 & \frac{27}{2} \\
 12 & 6 & 31 & \frac{27}{2} \\
 4 & \frac{81}{2} & \frac{81}{2} & \frac{115}{2} \\
\end{pmatrix}$} & 
$\begin{array}{l}G_1 \text{ in I2, II2} \\ 
\end{array}$
 \\\cline{2-2}\cline{5-5}
& 
{\footnotesize $\begin{array}{l}(\3, \2, \1, \pm \frac{1}{6}) + (\1, \2, \1, \pm\frac{1}{2}) \subset {\bf 16}_F, \\
(\1, \2, \2, 0)_1\subset {\bf 10}_H,\\
(\1, \2, \2, 0)_2 + (\1, \1, \3, \pm1) \subset \overline{\bf 126}_H, \\ 
(\1, \1, \1, 0)_1\subset {\bf 45}_H
\end{array}$} &
&
& 
$\begin{array}{l}G_2 \text{ in II4, III4} \\
G_3 \text{ in III7, IV1}
\end{array}$
 \\\cline{2-5}
& 
{\footnotesize $\begin{array}{l}(\3, \2, \1, \pm\frac{1}{6}) + (\1, \2, \1, \pm\frac{1}{2}) \subset {\bf 16}_F, \\
(\1, \2, \2, 0)_1\subset {\bf 10}_H,\\
(\1, \2, \2, 0)_2 + (\1, \1, \3, \pm1) \subset \overline{\bf 126}_H, \\
(\1, \1, \3, 0) + (\1, \3, \1, 0) \subset {\bf 45}_H 
\end{array}$} &
{\footnotesize $\begin{pmatrix} -7 \\ -\frac{4}{3} \\ -\frac{4}{3} \\ 7 \end{pmatrix}$} &
{\footnotesize $\begin{pmatrix}
 -26 & \frac{9}{2} & \frac{9}{2} & \frac{1}{2} \\
 12 & \frac{149}{3} & 6 & \frac{27}{2} \\
 12 & 6 & \frac{149}{3} & \frac{27}{2} \\
 4 & \frac{81}{2} & \frac{81}{2} & \frac{115}{2} \\
\end{pmatrix}$} &
$\begin{array}{l}G_2 \text{ in II9, III5} \end{array}$
 \\\hline
\multirow{4}{*}{$G_{421}$} & 
{\footnotesize $\begin{array}{l}(\4, \2, 0) + (\overline{\4}, \1, \pm\frac{1}{2}) \subset {\bf 16}_F, \\
(\1, \2, \frac{1}{2})\subset {\bf 10}_H, \\  
({\bf 15}, \2, \frac{1}{2}) + ({\bf 10}, \1, -1) \subset \overline{\bf 126}_H \end{array}$} &
$\begin{pmatrix} -7 \\ -\frac{2}{3} \\10 \end{pmatrix}$ &
$\begin{pmatrix}
 \frac{265}{2} & \frac{57}{2} & \frac{43}{2} \\
 \frac{285}{2} & \frac{115}{3} & 8 \\
 \frac{645}{2} & 24 & 51 \\
\end{pmatrix}$ &
$\begin{array}{l}G_1 \text{ in I3, II5,6,}\\ \text{III1}\end{array}$
\\\cline{2-5}
& 
{\footnotesize $\begin{array}{l}(\4, \2, 0) + (\overline{\4}, \1, \pm\frac{1}{2}) \subset {\bf 16}_F, \\
(\1, \2, \frac{1}{2})\subset {\bf 10}_H, \\  
({\bf 15}, \2, \frac{1}{2}) + ({\bf 10}, \1, -1) \subset \overline{\bf 126}_H, \\ 
({\bf 15}, \1, 0)\subset {\bf 45}_H \end{array}$} &
$\begin{pmatrix} -\frac{17}{3} \\ -\frac{2}{3} \\ 10 \end{pmatrix}$ &
$\begin{pmatrix}
 \frac{1243}{6} & \frac{57}{2} & \frac{43}{2} \\
 \frac{285}{2} & \frac{115}{3} & 8 \\
 \frac{645}{2} & 24 & 51 \\
\end{pmatrix}$ &
$\begin{array}{l}G_2 \text{ in II12,} \\ \text{III9,10, IV3}\end{array}$
 \\\hline
\multirow{8}{*}{$G_{422}$} & 
{\footnotesize $\begin{array}{l}(\4, \2, \1) + (\overline{\4}, \1, \2) \subset {\bf 16}_F, \\
(\1, \2, \2)\subset {\bf 10}_H, \\
({\bf 15}, \2, \2) + ({\bf 10}, \1, \3)\subset \overline{\bf 126}_H, \\  
(\1, \1, \3) \subset \overline{\bf 45}_H \end{array}$} &
$\begin{pmatrix} -\frac{7}{3} \\ 2 \\ \frac{28}{3} \end{pmatrix}$ &
$\begin{pmatrix}
 \frac{2435}{6} & \frac{105}{2} & \frac{249}{2} \\
 \frac{525}{2} & 73 & 48 \\
 \frac{1245}{2} & 48 & \frac{835}{3} \\
\end{pmatrix}$ & 
$\begin{array}{l} G_1 \text{ in I4, II7,} \\ 
G_2 \text{ in II5, III1} \end{array}$
\\\cline{2-5}
& 
{\footnotesize $\begin{array}{l}(\4, \2, \1) + (\overline{\4}, \1, \2) \subset {\bf 16}_F, \\
(\1, \2, \2)\subset {\bf 10}_H, \\
({\bf 15}, \2, \2) + ({\bf 10}, \1, \3)\subset \overline{\bf 126}_H, \\  
({\bf 15}, \1, \1) \subset \overline{\bf 45}_H \end{array}$} &
$\begin{pmatrix} -1 \\ 2 \\ \frac{26}{3} \end{pmatrix}$ &
$\begin{pmatrix}
 \frac{961}{2} & \frac{105}{2} & \frac{249}{2} \\
 \frac{525}{2} & 73 & 48 \\
 \frac{1245}{2} & 48 & \frac{779}{3} \\
\end{pmatrix}$ & 
$\begin{array}{l} G_2 \text{ in II1, III2} \end{array}$
\\\cline{2-5}
& 
{\footnotesize $\begin{array}{l}(\4, \2, \1) + (\overline{\4}, \1, \2) \subset {\bf 16}_F, \\
(\1, \2, \2)\subset {\bf 10}_H, \\
({\bf 15}, \2, \2) + ({\bf 10}, \1, \3)\subset \overline{\bf 126}_H, \\  
(\1, \1, \3) + ({\bf 15}, \1, \1) \subset \overline{\bf 45}_H \end{array}$} &
$\begin{pmatrix} -1 \\ 2 \\ \frac{28}{3} \end{pmatrix}$ &
$\begin{pmatrix}
 \frac{961}{2} & \frac{105}{2} & \frac{249}{2} \\
 \frac{525}{2} & 73 & 48 \\
 \frac{1245}{2} & 48 & \frac{835}{3} \\
\end{pmatrix}$ & 
$\begin{array}{l} G_2 \text{ in II10, III3} \\ 
G_3 \text{ in III8,10,} \\
\text{IV2,3} \\ 
\end{array}$
\\\hline
 \multicolumn{5}{|l|}{Continue on the next page}
\\\hline
\end{tabular}
\end{table}

\begin{table}[H]
\centering
\begin{tabular}{|p{10mm}|p{57mm}|c|c| p{25mm}|}
\hline
 \multicolumn{5}{|l|}{Table~\ref{tab:beta_coefficients} (Continued)}
\\\hline
\multirow{34}{*}{ $G_{422}^C$} & 
{\footnotesize $\begin{array}{l}(\4, \2, \1) + (\overline{\4}, \1, \2) \subset {\bf 16}_F, \\
(\1, \2, \2)\subset {\bf 10}_H, \\  
({\bf 15}, \2, \2) \!+\! ({\bf 10}, \1, \3)\! + \!(\overline{\bf 10}, \3, \1) \subset \overline{\bf 126}_H, \\
(\1, \3, \1) + (\1, \1, \3) \subset {\bf 45}_H \end{array}$} &
\multirow{4}{*}{$\begin{pmatrix} \frac{2}{3} \\ \frac{28}{3} \\ \frac{28}{3} \end{pmatrix}$} &
\multirow{4}{*}{$\begin{pmatrix}
 \frac{3551}{6} & \frac{249}{2} & \frac{249}{2} \\
 \frac{1245}{2} & \frac{835}{3} & 48 \\
 \frac{1245}{2} & 48 & \frac{835}{3} \\
\end{pmatrix}$} &
$\begin{array}{l}G_1 \text{ in I5}, \\ G_2 \text{ in II6}\end{array}$
\\\cline{2-2}\cline{5-5}
& 
{\footnotesize $\begin{array}{l}(\4, \2, \1) + (\overline{\4}, \1, \2) \subset {\bf 16}_F, \\
(\1, \2, \2)\subset {\bf 10}_H, \\  
({\bf 15}, \2, \2) \!+\! ({\bf 10}, \1, \3) \!+\! (\overline{\bf 10}, \3, \1) \subset \overline{\bf 126}_H, \\
(\1, \3, \1)+(\1, \1, \3) \subset {\bf 45}_H, \\ 
(\1, \1, \1) \subset {\bf 210}_H
\end{array}$} &
& &
$\begin{array}{l}G_2 \text{ in II7}, \\ 
G_3 \text{ in III1} \end{array}$
\\\cline{2-5}
& 
{\footnotesize $\begin{array}{l}(\4, \2, \1) + (\overline{\4}, \1, \2) \subset {\bf 16}_F, \\
(\1, \2, \2)\subset {\bf 10}_H, \\  
({\bf 15}, \2, \2) \!+\! ({\bf 10}, \1, \3) \!+\! (\overline{\bf 10}, \3, \1) \subset \overline{\bf 126}_H, \\
({\bf 15}, \1, \1) \subset {\bf 210}_H \end{array}$} &
&
&
$\begin{array}{l}G_2 \text{ in II2}\end{array}$
\\\cline{2-2} \cline{5-5}
& 
{\footnotesize $\begin{array}{l}(\4, \2, \1) + (\overline{\4}, \1, \2) \subset {\bf 16}_F, \\
(\1, \2, \2)\subset {\bf 10}_H, \\  
({\bf 15}, \2, \2) \!+\! ({\bf 10}, \1, \3) \!+\! (\overline{\bf 10}, \3, \1) \subset \overline{\bf 126}_H, \\
({\bf 15}, \1, \1) \subset {\bf 45}_H \end{array}$} &
$\begin{pmatrix} 2 \\ \frac{26}{3} \\ \frac{26}{3} \end{pmatrix}$ &
$\begin{pmatrix}
 \frac{1333}{2} & \frac{249}{2} & \frac{249}{2} \\
 \frac{1245}{2} & \frac{779}{3} & 48 \\
 \frac{1245}{2} & 48 & \frac{779}{3} \\
\end{pmatrix}$ &
$\begin{array}{l}G_2 \text{ in II3}\end{array}$
\\\cline{2-2}\cline{5-5}
& 
{\footnotesize $\begin{array}{l}(\4, \2, \1) + (\overline{\4}, \1, \2) \subset {\bf 16}_F, \\
(\1, \2, \2)\subset {\bf 10}_H, \\  
({\bf 15}, \2, \2) \!+\! ({\bf 10}, \1, \3) \!+\! (\overline{\bf 10}, \3, \1) \subset \overline{\bf 126}_H, \\
({\bf 15}, \1, \1) \subset {\bf 45}_H, \\ 
(\1, \1, \1) \subset {\bf 210}_H
\end{array}$} &
&
&
$\begin{array}{l}
G_3 \text{ in III2} \end{array}$
\\\cline{2-5}
& 
{\footnotesize $\begin{array}{l}(\4, \2, \1) + (\overline{\4}, \1, \2) \subset {\bf 16}_F, \\
(\1, \2, \2)\subset {\bf 10}_H, \\  
({\bf 15}, \2, \2) \!+\! ({\bf 10}, \1, \3) \!+\! (\overline{\bf 10}, \3, \1) \subset \overline{\bf 126}_H, \\
({\bf 15}, \3, \1)+({\bf 15}, \1, \3) \subset {\bf 210}_H
\end{array}$} &
$\begin{pmatrix} \frac{26}{3} \\ \frac{56}{3} \\ \frac{56}{3} \end{pmatrix}$ &
$\begin{pmatrix}
 \frac{6239}{6} & \frac{441}{2} & \frac{441}{2} \\
 \frac{2205}{2} & \frac{1619}{3} & 48 \\
 \frac{2205}{2} & 48 & \frac{1619}{3} \\
\end{pmatrix}$ &
$\begin{array}{l}G_2 \text{ in II11}, \\ G_3 \text{ in III3}\end{array}$
\\\cline{2-5}
& 
{\footnotesize $\begin{array}{l}(\4, \2, \1) + (\overline{\4}, \1, \2) \subset {\bf 16}_F, \\
(\1, \2, \2)\subset {\bf 10}_H, \\  
({\bf 15}, \2, \2) \!+\! ({\bf 10}, \1, \3) \!+\! (\overline{\bf 10}, \3, \1) \subset \overline{\bf 126}_H, \\
({\bf 15}, \1, \1) \subset {\bf 45}_H, \\ 
({\bf 15}, \1, \1) \subset {\bf 210}_H
\end{array}$} &
$\begin{pmatrix} \frac{10}{3} \\ \frac{26}{3} \\ \frac{26}{3} \end{pmatrix}$ &
$\begin{pmatrix}
 \frac{4447}{6} & \frac{249}{2} & \frac{249}{2} \\
 \frac{1245}{2} & \frac{779}{3} & 48 \\
 \frac{1245}{2} & 48 & \frac{779}{3} \\
\end{pmatrix}$ &
$\begin{array}{l}
G_3 \text{ in III4} \end{array}$
\\\cline{2-5}
& 
{\footnotesize $\begin{array}{l}(\4, \2, \1) + (\overline{\4}, \1, \2) \subset {\bf 16}_F, \\
(\1, \2, \2)\subset {\bf 10}_H, \\  
({\bf 15}, \2, \2) \!+\! ({\bf 10}, \1, \3) \!+\! (\overline{\bf 10}, \3, \1) \subset \overline{\bf 126}_H, \\
(\1, \3, \1)+(\1, \1, \3) \subset {\bf 45}_H, \\ 
({\bf 15}, \1, \1) \subset {\bf 210}_H
\end{array}$} &
&
&
$\begin{array}{l}
G_3 \text{ in III5} \end{array}$
\\\cline{2-2}\cline{5-5}
& 
{\footnotesize $\begin{array}{l}(\4, \2, \1) + (\overline{\4}, \1, \2) \subset {\bf 16}_F, \\
(\1, \2, \2)\subset {\bf 10}_H, \\  
({\bf 15}, \2, \2) \!+\! ({\bf 10}, \1, \3) \!+\! (\overline{\bf 10}, \3, \1) \subset \overline{\bf 126}_H, \\
({\bf 15}, \1, \1)+(\1, \3, \1)+(\1, \1, \3) \subset {\bf 45}_H 
\end{array}$} &
$\begin{pmatrix} 2 \\ \frac{28}{3} \\ \frac{28}{3} \end{pmatrix}$ &
$\begin{pmatrix}
 \frac{1333}{2} & \frac{249}{2} & \frac{249}{2} \\
 \frac{1245}{2} & \frac{835}{3} & 48 \\
 \frac{1245}{2} & 48 & \frac{835}{3} \\
\end{pmatrix}$ &
$\begin{array}{l}
G_3 \text{ in III6,9}, \\
G_4 \text{ in IV1} \end{array}$
\\\cline{2-2}\cline{5-5}
& 
{\footnotesize $\begin{array}{l}(\4, \2, \1) + (\overline{\4}, \1, \2) \subset {\bf 16}_F, \\
(\1, \2, \2)\subset {\bf 10}_H, \\  
({\bf 15}, \2, \2) \!+\! ({\bf 10}, \1, \3) \!+\! (\overline{\bf 10}, \3, \1) \subset \overline{\bf 126}_H, \\
({\bf 15}, \1, \1) + (\1, \3, \1) + (\1, \1, \3) \subset {\bf 45}_H, \\ 
(\1, \1, \1) \subset {\bf 210}_H \end{array}$} &
&
&
$\begin{array}{l} G_4 \text{ in IV2,3} \end{array}$
\\\hline
$G_{3211}$ & 
{\footnotesize $\begin{array}{l}(\3, \2, 0, \frac{1}{6}) + (\overline{\3}, \1, \pm\frac{1}{2}, -\frac{1}{6}) \\
+ (\1, \2, 0, -\frac{1}{2}) + (\1, \1, \pm \frac{1}{2}, -\frac{1}{2}) \subset {\bf 16}_F, \\
(\1, \2, \frac{1}{2}, 0)_1 \subset {\bf 10}_H,\\
(\1, \2, \frac{1}{2}, 0)_2 + (\1, \1, 1, -1) \subset \overline{\bf 126}_H \end{array}$} &
$\begin{pmatrix} -7 \\ -3 \\ \frac{14}{3} \\ \frac{9}{2} \end{pmatrix}$ &
$\begin{pmatrix}
 -26 & \frac{9}{2} & \frac{3}{2} & \frac{1}{2} \\
 12 & 8 & 1 & \frac{3}{2} \\
 12 & 3 & 8 & \frac{15}{2} \\
 4 & \frac{9}{2} & \frac{15}{2} & \frac{25}{2} \\
\end{pmatrix}$ &
$\begin{array}{l}G_1 \text{ in I6, II8-12,} \\ \text{III3,5-10, IV1-3}\end{array}$
\\\hline
\end{tabular}
\caption*{{\bf Table~\ref{tab:beta_coefficients}} (Continued). }
\end{table}
\appendix

\bibliographystyle{JHEP}
\bibliography{SO10}

\end{document}